\documentclass[twocolumn,trackchanges]{aastex7}

\usepackage{multirow}
\newcommand{\ergcm}[1]{$\times 10^{#1}$ ph cm$^{-2}$ s$^{-1}$}
\begin{document}

%\title{Template \aastex v7 Article with Examples\footnote{Footnotes can be added to titles}}

\title{Exploring Hard X-ray Properties of $\gamma$-ray Emitting Narrow Line Seyfert-I Galaxies \\through NuSTAR Observations}
%\titlerunning{Hard X-ray Properties of Fermi-detected NLSy1s} 

\author[0000-0002-9126-1817]{Suvas Chandra Chaudhary}
\affiliation{Department of Physics, University of the Free State, 205 Nelson Mandela Dr., Bloemfontein, 9300, South Africa.}
\affiliation{Inter-University Centre for Astronomy and Astrophysics, Pune, Maharashtra 411007, India}
\email{suvas0101phy@gmail.com}

\author[0000-0002-1173-7310]{Raj Prince}
\affiliation{Department of Physics, Institute of Science, Banaras Hindu University, Varanasi-221005, India.}
\email{priraj@bhu.ac.in}

\author[0000-0003-1873-7855]{Brian Van Soelen}
\affiliation{Department of Physics, University of the Free State, 205 Nelson Mandela Dr., Bloemfontein, 9300, South Africa.}
\email{VanSoelenB@ufs.ac.za}

\author[0000-0001-8890-5418]{Pieter Meintjes}
\affiliation{Department of Physics, University of the Free State, 205 Nelson Mandela Dr., Bloemfontein, 9300, South Africa.}
\email{MeintjPJ@ufs.ac.za}

\correspondingauthor{Raj Prince}
\email{priraj@bhu.ac.in}

\begin{abstract}

%With the launch of the Fermi-LAT observatory in 2008, more new gamma-ray objects were discovered, mostly dominated by blazars. In addition, some of the narrow line Seyfert 1 (NLSy1) galaxies were observed in gamma-rays, but in smaller numbers, making them different from other NLSy1 galaxies. We call them gamma-ray-detected NLSy1 galaxies, and they are believed to have strong relativistic jets similar to blazars and large viewing angles compared to blazars. Scientists still wonder how these objects are different with respect to blazars, which also happen to host a supermassive black hole at the center and possess a strong relativistic jet pointing toward the Earth within a few degrees. 
We studied the six gamma-ray-detected Narrow Line Seyfert 1 (NLSy1) galaxies using the hard X-ray observations from Nuclear Spectroscopic Telescope Array (NuSTAR) and optical g- \& r-band from Zwicky Transient Facility (ZTF). The X-ray spectra corresponding to all objects are well-fitted with a power-law spectral model, and a strong "redder-when-brighter" trend is seen, which is mostly seen in Blazars. The X-ray light curves were produced for all the available observations, and the F$_{var}$ is estimated. In 1H 0323+342, we found that F$_{var}$  lies between 9$\%$ to 22$\%$, suggesting significant variability in the source. Similarly, for PKS 2004-447, we found F$_{var}$  lies between 10$\%$ to 21$\%$. We see a strong X-ray and $\gamma$-ray spectral index correlation among these objects, suggesting that these are produced through a similar process. Comparing the X-ray spectral index with other class objects, we see that NLSy1 galaxies are similar to LBL and IBL types. We see a negative trend of X-ray flux with the $\gamma$-ray luminosity in these objects, suggesting an anti-correlation between them. A similar trend is seen between the X-ray flux, total jet power, and disk luminosity. The X-ray spectral index also shows a negative trend with total jet power and disk luminosity. The optical variability amplitude (in magnitude) lies between 0.90 to 2.32, and the fractional variability varies from 13\% to 40\%. The color-magnitude plot shows mostly the redder-when-brighter (RWB) trend, suggesting $\gamma$-NLSy1 are much closer to FSRQs than BL Lacs. Our results, overall, summarize how the various parameters in gamma-ray-detected NLSy1 are connected.

\end{abstract}



\section{INTRODUCTION}
Active galactic nuclei (AGN) are the center of active galaxies, which comprises a supermassive black hole and the accretion disk as main components, and in some of the AGN, bi-polar jets have also been observed perpendicular to the disk plane. These AGN are randomly oriented in the Universe, and their central part is completely covered with a dusty molecular torus \citep{1983Natur.303..158A}. %The spectroscopic observations show broad and narrow emission lines in their optical spectra. %depending upon their viewing angle, and they are classified as broad- or narrow-line galaxies. 
The Narrow-Line Seyfert-1 (NLSy1) galaxies are a subclass of AGN with the presence of narrow 'broad permitted lines' in their optical spectra. The Full Width Half Maximum (FWHM) of the brightest Balmer line, i.e., H$\beta$, is less than $\sim$ 2000 km/s. It has also been found that the NLSy1 shows a weak forbidden [OIII]$\lambda$5007 line with a flux ratio of  [OIII]/H$\beta$ $<$3 \citep{1985ApJ...297..166O, 1989ApJ...342..224G}. In addition, they also show strong FeII emission in their optical spectra, which typically anti-correlates with the O[III] emission and with the width of the broad Balmer emission lines. They also show interesting properties such as a strong soft excess and high amplitude variability in X-rays. Studies have shown that the NLSy1 carries the supermassive black holes (SMBHs) mass of 10$^6$ - 10$^8$ M$_{\odot}$ and a high accretion rate. The recent catalog created by \cite{2017ApJS..229...39R} has identified 11,101 galaxies as NLSy1 from the SDSS survey. On the other hand, the gamma-ray NLSy1 is very small in number and was mostly detected when the Fermi-LAT started operating in 2008. Gamma-ray detections have given new directions for studying these objects since they belong to the AGN class with strong jet emission. However, it is still unclear if they are similar to blazars or not, since their gamma-ray and X-ray properties differ greatly. Based on radio properties \citet{2016A&A...591A..98B} have shown that the  NLSy1s are the low-mass tail of the quasar distribution. 
 
Our study aims to explore the possibility of providing the parameter estimates of these objects and comparing them with blazar types such as flat-spectrum radio quasars (FSRQs) and BL Lacertae objects. This will help to determine if these objects are similar to blazars or have distinct properties. We chose the hard X-ray observations to achieve our aim, as they have been extensively observed in NuSTAR. 
Some of the sources from our sample have already been studied in \cite{2019A&A...632A.120B}, where they have used Swift-XRT, XMM-Newton, and NuSTAR observations to determine the origin of these emissions. Their result suggests the X-ray emission in NLSy1 can be contributed from both the jet and the accretion disk corona. We include more observations of these sources, along with two additional objects, to derive the more general properties of these objects. Additionally, we also incorporate optical data from ZTF to understand the optical behaviour of these objects.  

We aim to establish some correlation between the hard X-ray temporal and spectral properties of these AGNs with known FSRQs and BL Lacs; this will help to probe the fundamental difference between these classes.
As studies suggest, blazars are more powerful objects in gamma rays and have strong relativistic jets with power ranging between 10$^{44}$-10$^{47}$ ergs/s. The jet power estimated in \citet{2019ApJ...872..169P} reveals a similar power as a blazar, but these objects differ very much in gamma-ray and X-ray properties.
%It would be interesting to see if gamma-ray-detected NLSy1 has similar jet power or not.
This will pinpoint the main fundamental question that we aim to answer, which is, at the fundamental level, how NLSy1s are different from blazars. The broadband SED modeling performed by \citet{2019ApJ...872..169P} for a sample of NLSy1 galaxies shows that the accretion disk mostly dominates the optical-UV part of the spectrum, and the gamma-ray emission is from the Jet. Therefore, these sources are the best candidates to study the disk-jet coupling to understand how the disk and jet are connected.

In section \ref{sample}, we discuss the sample selection and its properties, followed by the NuSTAR analysis in section \ref{data}. In section \ref{analysis}, we discuss various methodologies used in this study. In section \ref{discussion}, we discuss the derived temporal and spectral parameters in context with FSRQs and BL Lacs, and in section, we present the results and discuss the work in section \ref{conclusion} presents the conclusion of the work.\\

\section{SAMPLE SELECTION}  \label{sample}
In this study, we have made a sample of NLSy1 emitting gamma rays with {\it NuSTAR} data. With the literature survey \citep{2018MNRAS.481.5046R, 2022Univ....8..587F, 2024MNRAS.527.7055P}, we have found six $\gamma$-ray detected NLSy1, with 14 epochs. The basic information of the selected sources is listed in the Table \ref{tab: Sample_info}.

\subsection{1H 0323+342:}
1H 0323+342 is the closest NLSy1 galaxy in our sample ($z=0.06$, \cite{2007ApJ...658L..13Z}). \cite{2016RAA....16..176F} analyzed the VLBA images at 15 GHz and predicted superluminal motion with an apparent speed of approximately $1-7 c$ while also constraining the jet viewing angle between $4$ and $13$ degrees. The observed blackhole mass measured in literature is around $\sim 10^{6-7} M_{\odot}$ \citep{2015AJ....150...23Y,2017MNRAS.464.2565L,2016ApJ...824..149W}. X-ray timing and spectral analysis reported in \citep{2020MNRAS.496.2922M,2018ApJ...866...69P,2018IJMPD..2744001B} suggest a soft excess below 2 keV and hard excess above 35 keV. The above studies also hint at the X-ray emission from the combination of the hot corona around the central engine and the jet. The source showed a multiwavelength flare during 2013, suggesting \citep{2018IJMPD..2744001B} three times the flux enhancement in X-ray and eighteen times the flux increase in $\gamma$-ray. Radio, Infrared, X-ray, and $\gamma$-ray properties presented in \citep{2018rnls.confE..16Y} suggest a correlated long-term variability in a high flux state. X-ray spectral properties discussed in \citep{2004Sci...306..998G,2009AdSpR..43..889F,2012nsgq.confE..10F,2019Univ....5..199F} were explained by disc-jet coupling during the low jet state. They have also found the Compton-dominated disk corona emission ($\Gamma_{0.3-10} \sim 2$), while the hard X-ray tail ($\Gamma_{3-10} \sim 1.4$) was observed during the high jet activity. On top of that, SWIFT-BAT has also detected a high flux and hard spectrum, while the INTEGRAL/IBIS monitoring reported a soft spectrum in the low state \citep{2009AdSpR..43..889F}.

\subsection{PKS 2004-447:}
PKS 2004-447 (z = 0.24), located in the southern hemisphere, shows a characteristic of the spectral energy distribution of beamed, jet-dominated AGN \citep{2006MNRAS.370..245G}. Its radio morphology indicates a line-of-sight angle of less than \( 50^\circ \) \citep{2016A&A...588A.146S,2015MNRAS.453.4037O}, which is significantly higher compared to typical blazar sources. An X-ray analysis of this source in the energy range of 0.5–10 keV, based on observations of XMM-Newton and Swift \citep{2013EPJWC..6104017K,2016A&A...585A..91K}, indicates variability on timescales of months to years. In particular, the X-ray luminosity spans two orders of magnitude, ranging from $10^{44} erg s^{-1}$ to $10^{46} erg s^{-1}$, making this source a particularly intriguing NLSy1 in the X-ray band. Several studies have been conducted to estimate the mass of the central black hole, which turns out to be $10^{7-8} M_{\odot}$ \citep[e.g.,][]{2015A&A...575A..13F,2016ApJ...832..157K,2016MNRAS.458L..69B}. The source does not exhibit significant variability, unlike other $\gamma$-NLSy1. Its X-ray spectra can be explained by a single power-law model, indicating a prominent jet contribution similar to blazars \citep{2016A&A...585A..91K, 2021A&A...654A.125B}. \cite{2006MNRAS.370..245G} have reported the presence of a weak soft excess, below 1 keV, during the 2004 epoch (high flux state) of XMM observations. However, the new XMM observations presented in \citep{2016A&A...585A..91K, 2020Univ....6..136F} lack the presence of the soft excess. However, photon index \& unabsorbed flux measurements show that PKS 2004-447 has similar features as compact steep-spectrum (CSS) sources. \cite{2020Univ....6..136F} argued that one would anticipate that the soft excess will appear when the jet continuum shrinks if the accretion disk causes it.

\subsection{PKS 1502+036:}
The PKS 1502+036 was discovered by Fermi-LAT \citep{2009ApJ...707L.142A, 2012arXiv1205.0402O} as a faint gamma-ray source with gamma-ray luminosity, $L_{\gamma} \sim 10^{46}$. The source has been monitored using various ground- and space-based telescopes over a wide range of frequencies. \cite{2023ApJ...942...51W,2019ApJ...872..169P,2016MNRAS.463.4469D} studied X-ray emission and $\gamma$-ray emission. In addition, the broad emission lines in its optical spectrum indicate that it harbors an SMBH at its center with a mass of $10^{8} M_{\odot}$\citep{2016MNRAS.463.4469D,2017ApJ...842...96R}. \cite{2019ApJ...884...15S} analyzed long-term radio, optical, and $\gamma$-ray data and carried out a cross-correlation function to study time delay in photon emission at various wavelengths and pinpointed the $\gamma$-ray and radio-emitting regions within the jet.
A multi-wavelength study published in \citep{2023MNRAS.523..441Y} suggests low- and high-flux states; the optical/UV variability is about 2$\%$, while the X-ray variability can be as high as 47$\%$. Fermi-LAT has also detected the GeV flare in PKS 1502+036 like few other $\gamma$-NLSy1 such as 1H 0323+342, and PMN J0948+0022, with a flux seventeen times its average \citep{2015ATel.8447....1D}, with spectral index of $\Gamma_{0.1-300 GeV} = 2.57\pm0.17$ and flux value (3.9$\pm$1.52)\ergcm{-6} \citep{2016ApJ...820...52P}. The broadband SED in \citep{2016MNRAS.463.4469D, 2016ApJ...820...52P} shows that the high energy photons, X-ray to $\gamma$-ray in SED can be successfully fitted by inverse-Compton scattering of torus thermal photons, while synchrotron and disk accretion can replicate the optical-UV spectrum and concluded that the broadband SED, high-energy emission mechanism, and $\gamma$-ray activity are comparable to those of FSRQs.
%{\bf{\cite{2010ApJ...710..810A} have reported a three times flux enhancement within 12 hours using Fermi-LAT. A broadband flare was reported in \citep{2016A&A...590A..48K, 2010ApJ...710..810A}, suggesting that the seed photons for high-energy photons mainly originate outside the broad line region (BLR)}}. Strong relativistic beaming effects are suggested by the flare episodes observed in PKS 1502+036. X-ray spectral investigation reveals different features in PKS 1502+036. The spectrum shows a soft excess component and a broken power-law component with photon indices $\Gamma_{1} \sim2.10$ and $\Gamma_{2} \sim1.52$, and break energy at $E_{break} \sim 0.62$ keV \cite{2014xru..confE..56D}. 

\subsection{RGB J1644+263:}
RGB J1644+2619 was detected by Fermi-LAT in 2015, with a average $L_{\gamma} \sim 10^{44} erg/s$ and $\Gamma_{\gamma}=2.79$ \citep{galaxies4030011}. It exhibited multiple flares from 2009 to 2025, with the highest flux reaching nine times the average flux \citep{2018MNRAS.476...43L}. The gamma-ray properties of this source align with those of typical jetted sources, including NLSy1 and blazars. Radio observations suggest the presence of a high radio loudness ($R \sim 250$) and a flat radio spectrum and core-dominant one-sided jet, with a jet speed $\beta=0.983$ and viewing angle $\theta<5\deg$ similar to other NLSy1 galaxies \citep{Doi_2011, Doi_2012,2016PASJ...68...73D}. X-ray analysis reported in \citep{2011nlsg.confE..24F, Yuan_2008,2018MNRAS.476...43L} exhibits a soft excess in lower energy and a hard X-ray emission above 2 keV, which is a typical behavior of NLSy1. The source shows significant variability in optical/UV with a variability amplitude of $\sim$1.4-18 to X-rays $\sim$2.7, over a timescale of days to months. These results suggest the dominance of jet emissions over the disk emission.

\subsection{PMN J0948+0022:}
PMN J0948+0022 (z = 0.5846) stands out as the first NLSy1 galaxy detected by Fermi-LAT \citep{2009ApJ...699..976A, 2009ApJ...707L.142A}, marking a significant discovery in the study of gamma-ray-emitting AGNs, known for its brightness across a wide range of wavelengths; this galaxy exhibits unique spectral properties that firmly categorize it as an NLSy1. Specifically, its optical spectrum displays relatively narrow H$\beta$ emission lines with FWHMH$\beta$ of approximately 1500 km s$^{-1}$, along with weak forbidden lines. Additionally, PMN J0948+0022 displays significant radio loudness and noticeable variability within its compact radio core, both of which are indicative of a relativistic jet. Observations of the galaxy’s emission from radio to gamma-ray wavelengths \citep{2012A&A...548A.106F} provide further insight into its central black hole properties. The black hole is estimated to have a mass around \(10^8 M_{\odot}\), and its accretion disk emits a luminosity of about 40$\%$ of the Eddington limit, signifying an energetic and efficient accretion process. Additionally, this system is observed at a very small viewing angle, meaning that the jet is likely oriented close to our line of sight, enhancing its apparent brightness because of relativistic effects. \cite{2023MNRAS.523..441Y} suggested that the observed X-ray emission is jet-dominated, while optical / UV contributes most strongly to disk processes. \cite{2021RNAAS...5..109M} studied the mid-infrared variability using {\it WISE} data, showing a significant bluer trend when brighter, similar to X-ray emitting blazars \citep[e. g., see][] {2018A&A...619A..93B,bhatta2024probingxraytimingspectral}.

\subsection{CGRaBS J1222+0413:}
CGRaBS J1222+0413 is the farthest NLSy1 in our sample located at redshifts $z=0.97$ \citep{10.1093/mnrasl/slv119}, first reported in \cite{2015MNRAS.454L..16Y}. The authors have also measured the mass of the central SMBH, which is \( M_{BH}\sim 10^8 M_{\odot}\).
Studies presented in \citep[e. g., see][]{2019MNRAS.483.3036O, Ackermann_2015} indicates high radio loudness, one-sided jet \citep{2016AJ....152...12L}, flat hard X-ray spectra with a power-law spectral index of $\Gamma\sim1.3$ and a bulk Lorentz factor of $\sim30$. Unlike other $\gamma$-NLSy1, this source shows twice disc emission over jet emission, which makes it an interesting candidate to investigate disc-jet coupling \citep{2019MNRAS.487..181K}. An extensive broadband study of this source from radio data (Effelsberg, Planck, FIRST, telescope), IR data (Herschel, Spitzer, WISE, 2MASS and VLT X-shooter) Optical/UV data (VLT X-shooter, SDSS, SWIFT-UVOT, XMM–Newton OM, GALEX and HST), X-ray data (ROSAT, XMM-Newton, Swift-XRT, NuSTAR and Swift-BAT) to $\gamma-ray$ data (Fermi-LAT) presented in \citep{2019MNRAS.487..181K} suggests a wide range of observed luminosity $\sim 10^{43 - 47} erg s^{-1}$. The above analysis also suggests a hard X-ray photon index $\Gamma \sim 1.5$ in X-ray data, suggesting a jet contribution in X-ray emission.

\begin{table*}[h]
    \centering
    \begin{tabular}{cccccccccc}\hline
        Name & 4FGL Name & RA(J2000) & Dec.(J2000)  &z  & $F_{1.4 GHz}$  & $d_L$ & $L_{\gamma}$ & $M_{BH}$ & $P_{jet}$\\
        &  &  &  & & mJy & Mpc& erg$s^{-1}$ &$M_{\odot}$ & erg$s^{-1}$  \\ \hline
       1H 0323+342 & J0324.8+3412 & 51.1715  & +34.1794 & 0.06 &  613.5 & 270.3 & $2.1 \times 10^{44}$ & 7.30 & $6.61 \times 10^{45}$ \\
       PKS 2004-447  & J2007.9-4432 & 301.9799 & -44.5789 & 0.24 & 471.0 & 1213.0 & $1.7 \times 10^{45}$ &  6.70 & $1.70 \times 10^{45}$\\
       %PKS 1441+25 & J1443.9+2501 & 220.9870 & +25.0290 &  0.94 & - & 6205.9  & $3.50\times10^{47}$  &7.83 \\
       PKS 1502+036  & J1505.0+0326 & 226.2769 & +03.4418 &  0.41 & 394.8 &2258.5 & $1.0 \times 10^{46}$ &7.60  & $ 12.30\times 10^{45}$ \\
       RGB J1644+263 & J1644.9+2620  & 251.1772 & +26.3203  & 0.14 & 128.4 & 666.2 & $2.7 \times 10^{44}$ &7.70 & $ 8.13\times 10^{45}$\\
       PMN J0948+0022 & J0948.9+0022  & 147.2388 & +00.3737 & 0.58 & 69.5 & 3426.3  & $7.5 \times 10^{46}$ & 8.18 & $ 128.82\times 10^{45}$ \\
      
       CGRaBS J1222+0413  & J1222.5+0414  & 185.5939 & +04.2210 & 0.97 & 800.3 & 6452.6 & $2.3 \times 10^{47}$ &  8.85 & $389.04 \times 10^{45}$\\
 \hline
    \end{tabular}
    \caption{Basic information about the sources in this work. Various source parameters such as {BH masses \citep{2016ApJ...824..149W, 2007ApJ...658L..13Z, 2017ApJS..229...39R, 2013MNRAS.431..210C, 2015AJ....150...23Y, 2001ApJ...558..578O}}, $L_{\gamma}$ \citep{2020MNRAS.496.2213D, 2022ApJS..260...53A}, radio flux at 1.4GHz \citep{1998AJ....115.1693C}, redshift, Disk luminosity \& Jet Power \citep{2019ApJ...872..169P,2014MNRAS.441.3375X}.}
    \label{tab: Sample_info}
\end{table*}

\section{NuSTAR OBSERVATIONS AND DATA REDUCTION} \label{data}
The Nuclear Spectroscopic Telescope Array (NuSTAR) is a high-energy X-ray satellite operating in the 3–79 keV energy range. It has an angular resolution of $18^{\prime\prime}$ (FWHM), a temporal resolution of 2$\mu$s, and an energy resolution of 0.4 keV at 6 keV and 0.9 keV at 60 keV (FWHM), making it an excellent instrument for hard X-ray imaging, timing, and spectral analysis. The telescope is equipped with two detector units, FPMA and FPMB, specifically designed to image astrophysical objects in hard X-rays \cite{2013ApJ...770..103H}. We have processed the raw data of all six sources with all 14 epochs. Our sample sources have exposures of 25-190 ks. Raw data products were initially processed using the NuSTAR Data Analysis Software (NuSTARDAS) package, {\it V0.4.9}. For data reduction and analysis, we used {\it HEASOFT V6.34} and {\it CALDB V20220413}. Calibrated and cleaned event files were generated using the standard {\it nupipeline} script. We extracted the source flux and spectra from a circular region with a $60^{\prime\prime}$ radius centered on the source location. We selected a nearby region with a $120^{\prime\prime}$ radius for background extraction, far enough from the source to avoid contamination. Light curves were created with a time bin of 20 minutes, and the spectra were re-binned using the {\it grppha} task to achieve a minimum of 20 counts per channel. We performed the spectral fitting using {\it XSPEC V12.12.1}\footnote{\url{https://heasarc.gsfc.nasa.gov/xanadu/xspec/}}\citep{1996ASPC..101...17A}. Table \ref{tab: NuSATAR_info} presents the X-ray spectral fitting results.

\begin{table*}
    \centering
    \begin{tabular}{cccccccccc} \hline
      Object   & Obs. Date & Obs. ID & $F_{var}$ & Time & $\Gamma$ & Flux &  $\chi^2_{r}$ & $N_H$ ($cm^{-2})$\\ \hline
       
       1H 0323+342  & 2014-03-15  & 60061360002 & 0.14$\pm$0.01 & 101.63 & 1.83$\pm$0.01 & 23.12$\pm$0.25 & 813.43/839 & $11.70 $\\ 
                    & 2018-08-14 & 60402003002 & 0.19$\pm$0.02 & 36.39 & 1.80$\pm$0.03 & 20.16$\pm$0.34 & 483.17/500 & -\\
                    & 2018-09-05 & 60402003010 & 0.15$\pm$0.02 & 30.42 & 1.79$\pm$0.02 & 20.08$\pm$0.41 & 446.36/446 &- \\
                    & 2018-08-18 & 60402003004  & 0.10$\pm$0.03 & 29.73 & 1.85$\pm$0.04 & 15.02$\pm$0.33 & 393.06/345 & -\\
                    & 2018-09-09 & 60402003012  &0.22$\pm$0.02 & 27.79 & 1.75$\pm$0.03 & 14.08$\pm$0.24 & 279.81/264 & -\\
                    & 2018-08-20 & 60402003006  & 0.11$\pm$0.02 & 26.40 & 1.85$\pm$0.02 & 30.78$\pm$0.48 & 481.55/484 & -\\
                    & 2018-08-24 & 60402003008  & 0.09$\pm$0.03 & 25.56 & 1.82$\pm$0.02 & 17.96$\pm$0.43 & 351.49/318 & -\\ 
                   & 2025-09-25 & 91101637002  & 0.12$\pm$0.07 & 48.09 & 1.83$\pm$0.01 & 72.07$\pm$0.70 & 832.69/843 & -\\ \\
                    
        PKS 2004-447  & 2016-10-23 & 60201045002 & 0.10$\pm$0.15 & 60.59 & 1.62$\pm$0.07 & 3.56$\pm$0.11 & 156.30/177 & $2.97$\\
                      & 2016-05-09 & 80201024002  & 0.16$\pm$0.09 & 49.21 & 1.66$\pm$0.10 & 3.09$\pm$0.12 & 132.22/132 & -\\
                      & 2019-11-01 & 90501649002  & 0.21$\pm$0.08 & 30.07 & 1.36$\pm$0.05 & 5.73$\pm$0.30 & 95.98/113 & -\\ \\
         PKS 1502+036  & 2017-02-12 & 60201044002 & 0.16$\pm$0.09 & 115.45 & 1.23$\pm$0.06 & 1.98$\pm$0.14 & 235.72/205 &$ 3.47 $ \\ \\
         RGB J1644+263 & 2018-01-18 & 60301017002 & 0.13$\pm$0.15 & 52.48 & 1.75$\pm$0.07 & 2.24$\pm$0.14 & 108.38/105 & $ 5.02$\\ \\
         PMN J0948+0022 & 2016-11-04 & 60201052002 & 0.10$\pm$0.01 & 192.69 & 1.41$\pm$0.02 & 6.89$\pm$0.11 & 628.57/668 & $ 4.73$ \\ \\
        CGRaBS J1222+0413  & 2017-06-27 & 60301018002  & 0.16$\pm$0.05 & 32.22 & 1.45$\pm$0.04 & 7.50$\pm$0.33 & 147.45/149 & $1.64$ \\  \hline
    \end{tabular}\\
    \caption{NuSTAR observations of the selected Gamma-ray emitting NLS1 sources: Col 1: Name; Col 2: NuSTAR observation date; Col 3: Observation ID;  Col 4: Fractional Variability; Col 5: Exposure Time (ks); Col 6: Photon Index; Col 7: Unabsorbed X-ray Flux ($10^{-12}erg s^{-1} cm^{-2}$) in 3.0-79.0 keV; Col 8: Reduced Chi-square statistic; Col 9: Column density ($10^{20} cm^{-2}$).}
    \label{tab: NuSATAR_info}
\end{table*}

\section{ANALYSIS METHODS} \label{analysis}
We cross-matched the Fermi-detected NLSy1 objects with the NuSTAR archive \citep{2018MNRAS.481.5046R, 2022Univ....8..587F}. We found a total of six objects, which were extensively observed with NuSTAR over the period. We produced the light curve as well as the spectrum for all the observation IDs available for all those objects. The basic information about the objects is tabulated in Table \ref{tab: Sample_info}. We modeled the spectrum with a simple absorbed power law for the full energy range (3-79 keV). The best-fit parameters, along with the reduced $\chi^2$, are presented in Table \ref{tab: NuSATAR_info}. The reduced $\chi^2$ suggests that a single power law can well explain the spectrum, which is mostly the case in blazars.

\subsection{FLUX VARIABILITY}
We have used Fractional Variability ($F_{var}$) as discussed in \citep{1990ApJ...359...86E, 1997ApJS..110....9R, Vaughan_2003}, to measure the variability in the light curve, for which we have binned it in 20 minutes. The $F_{var}$ can be estimated as:

\begin{equation} \label{FVs}
    F_{\rm var} = \sqrt{\frac{S^{2} - \bar{\sigma}^{2}_{\rm err}}{\bar{X}^2}},
\end{equation}
Where $S^2$ is the light curve variance, $\bar {\sigma^2}$ is the flux mean square error, and $\bar X$ is the mean flux. The associated error in the $F_{var}$ is obtained using,
\begin{equation}
   \sigma_{F_{\mathrm{var}}} =  \sqrt{\left(\frac{1}{\sqrt{2N}}\frac{\bar{\sigma}^{2}_{err}}{F_{var}}\frac{1}{\bar{X}^2}\right)^2 + \left(\sqrt{\frac{\bar{\sigma}^{2}_{err}}{N}}\frac{1}{\bar{X}^2}\right)^2}. 
\end{equation}

\begin{figure*}
    \centering

     \includegraphics[angle=0, width=0.480\linewidth]{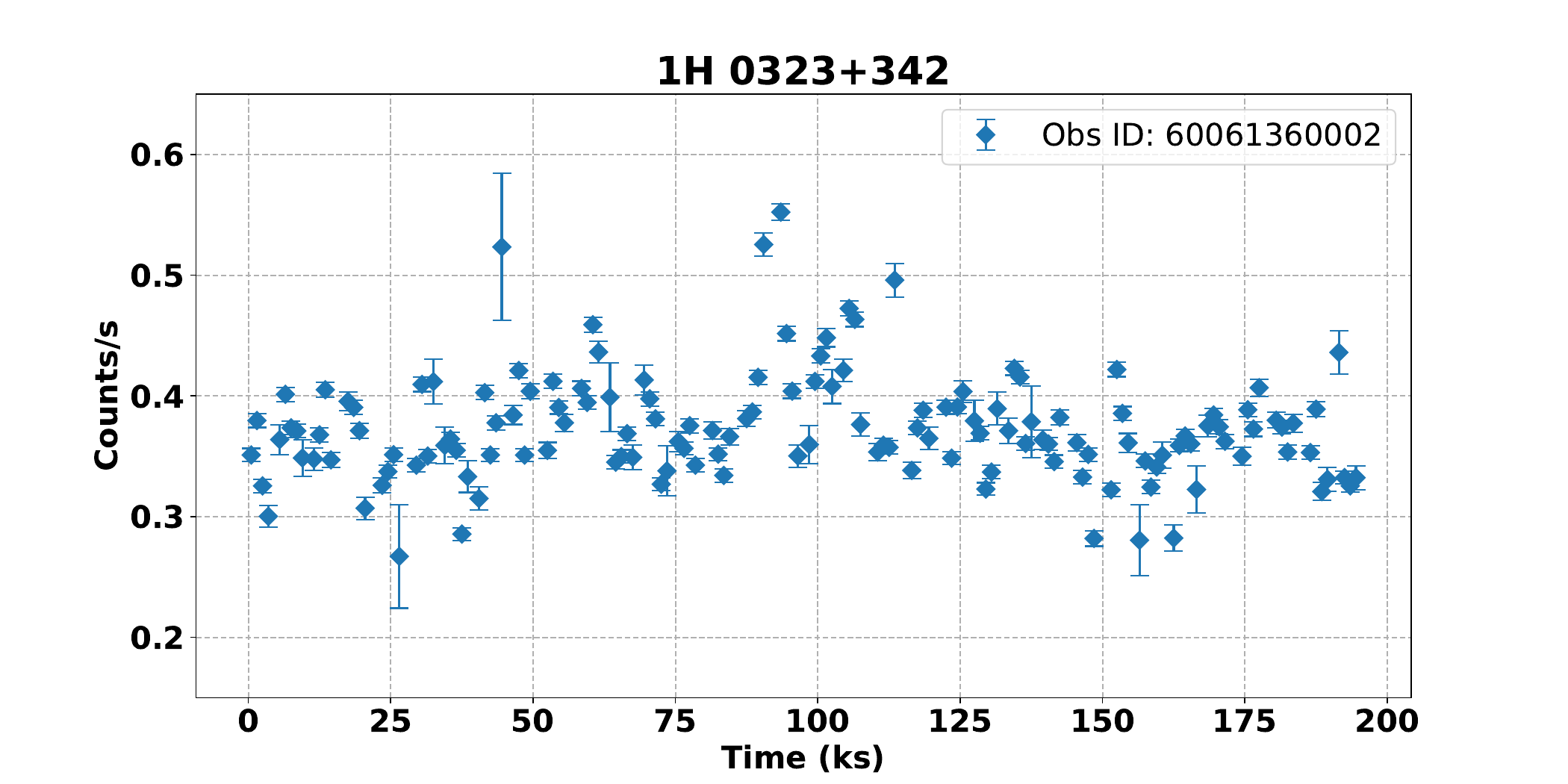}
   \includegraphics[angle=0, width=0.480\linewidth]{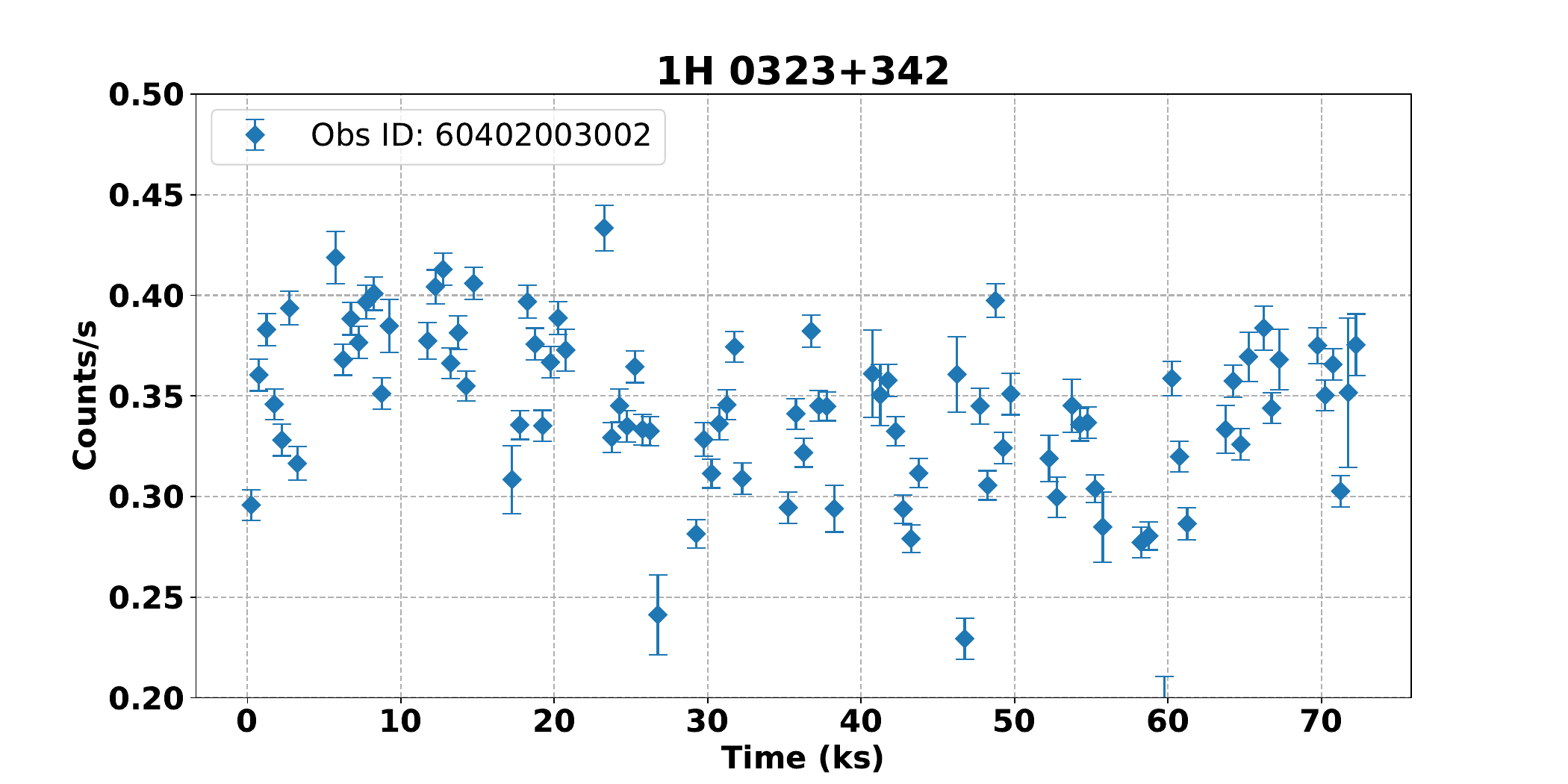}
   \includegraphics[angle=0, width=0.480\linewidth]{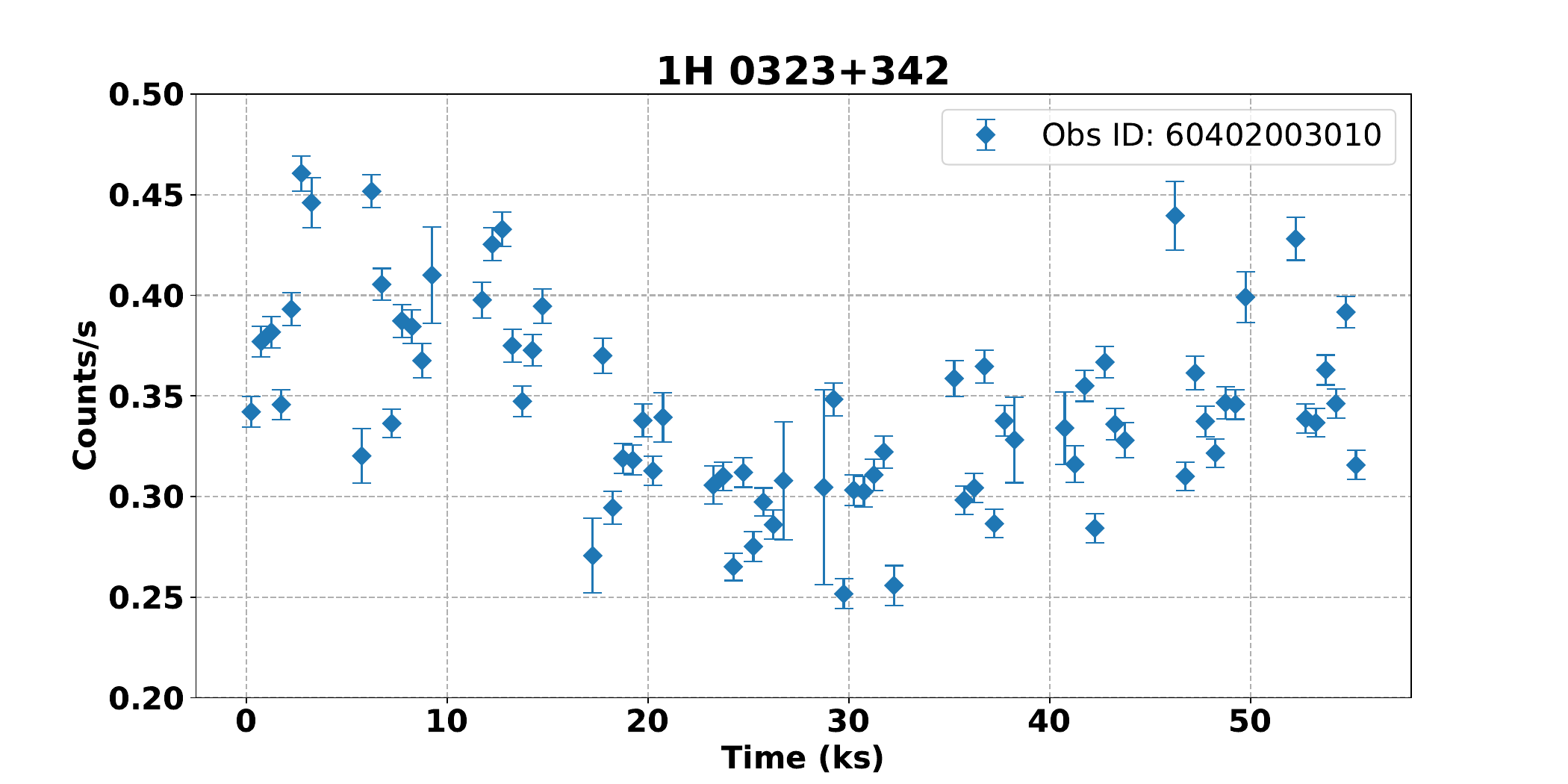}
    \includegraphics[angle=0, width=0.480\linewidth]{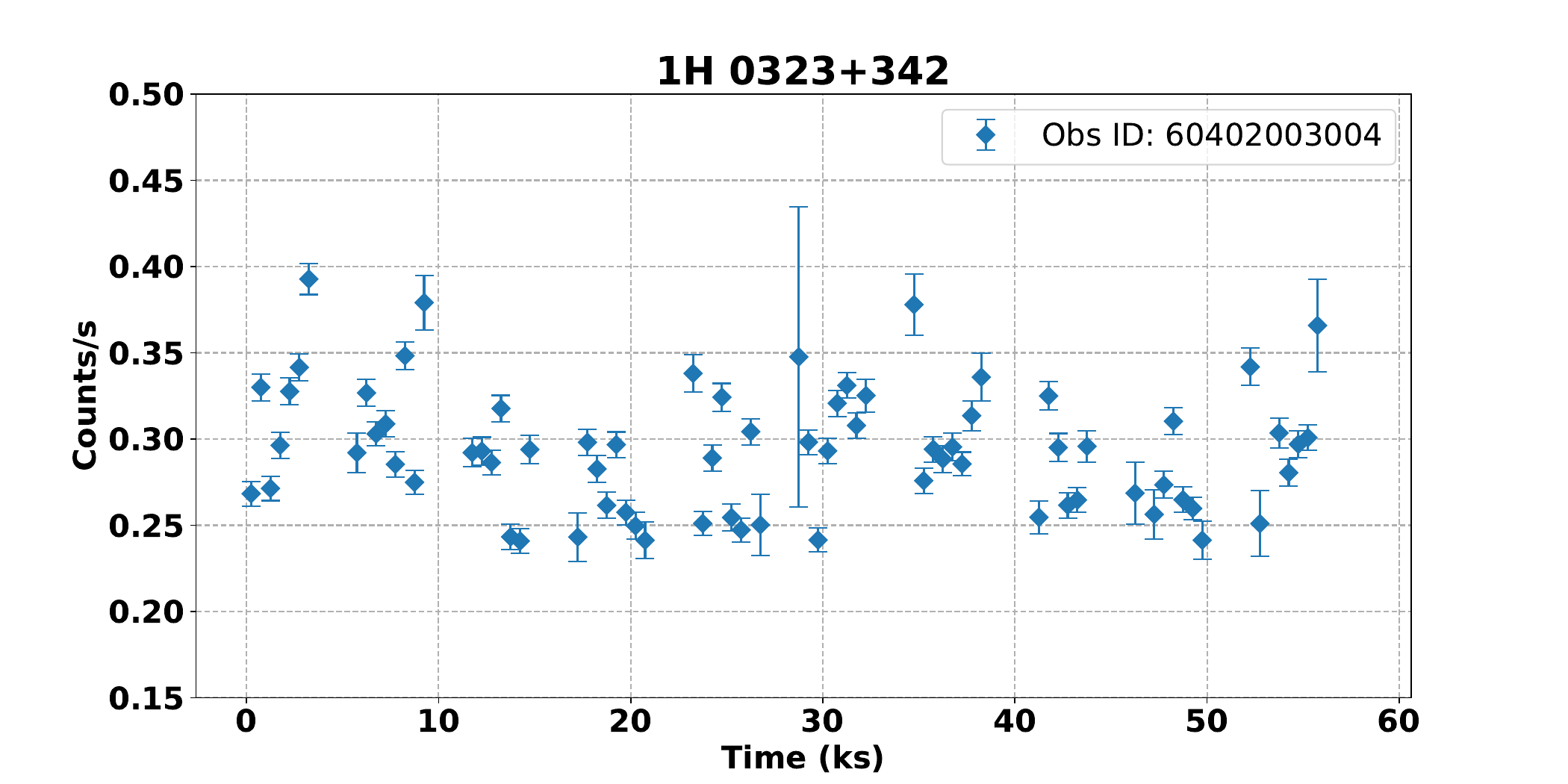}
     \includegraphics[angle=0, width=0.480\linewidth]{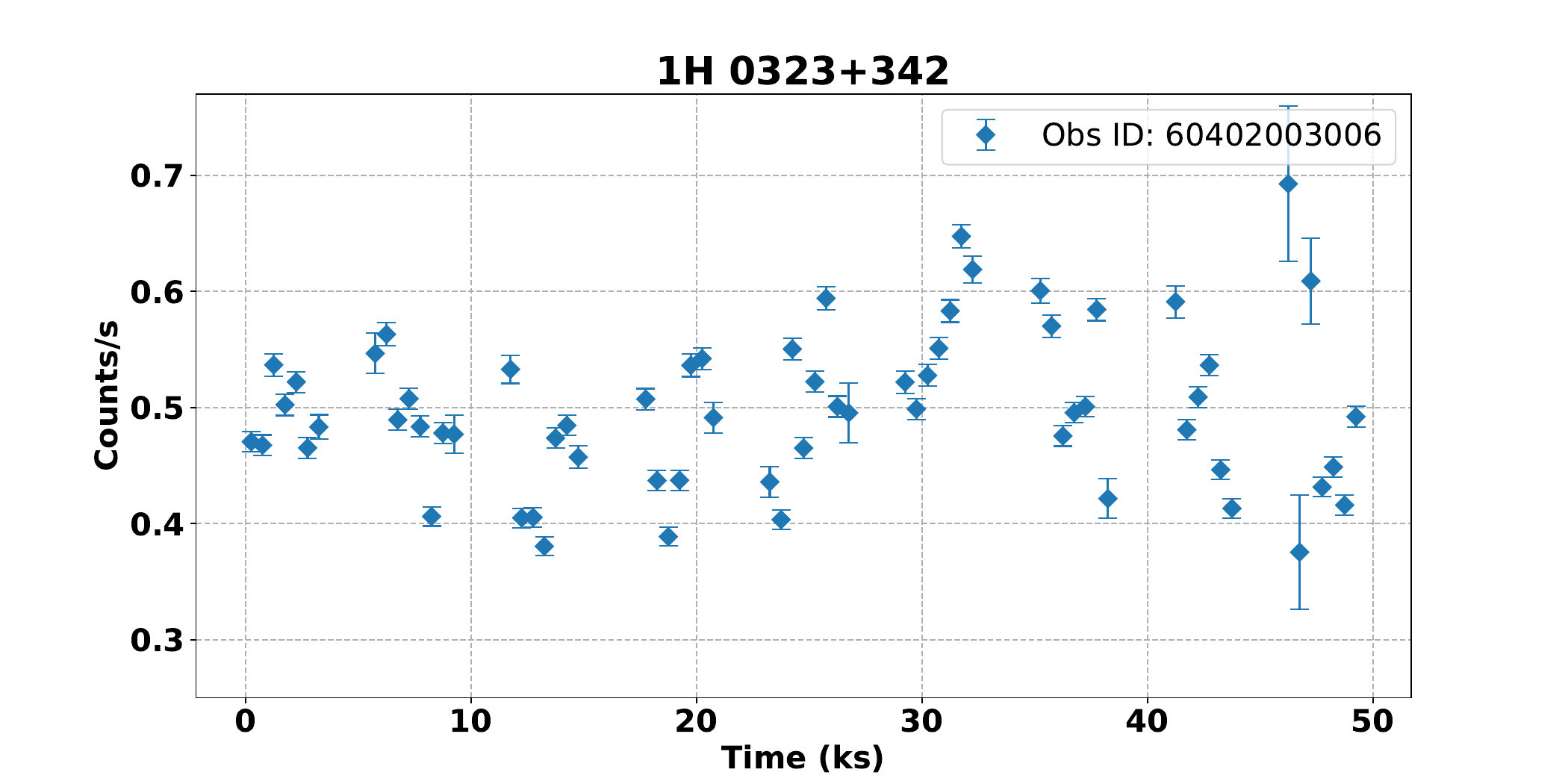}
     \includegraphics[angle=0, width=0.480\linewidth]{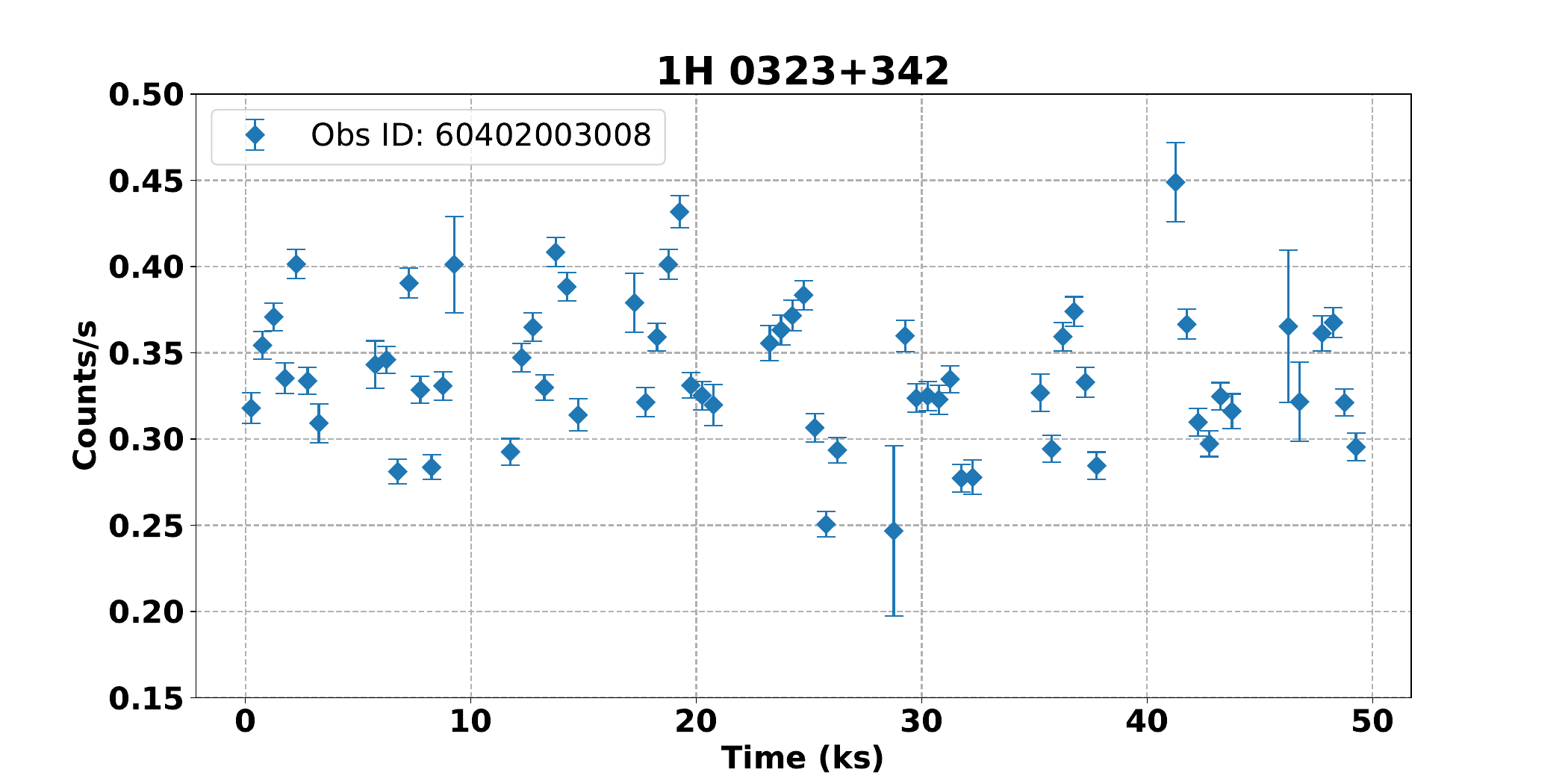}
      \includegraphics[angle=0, width=0.480\linewidth]{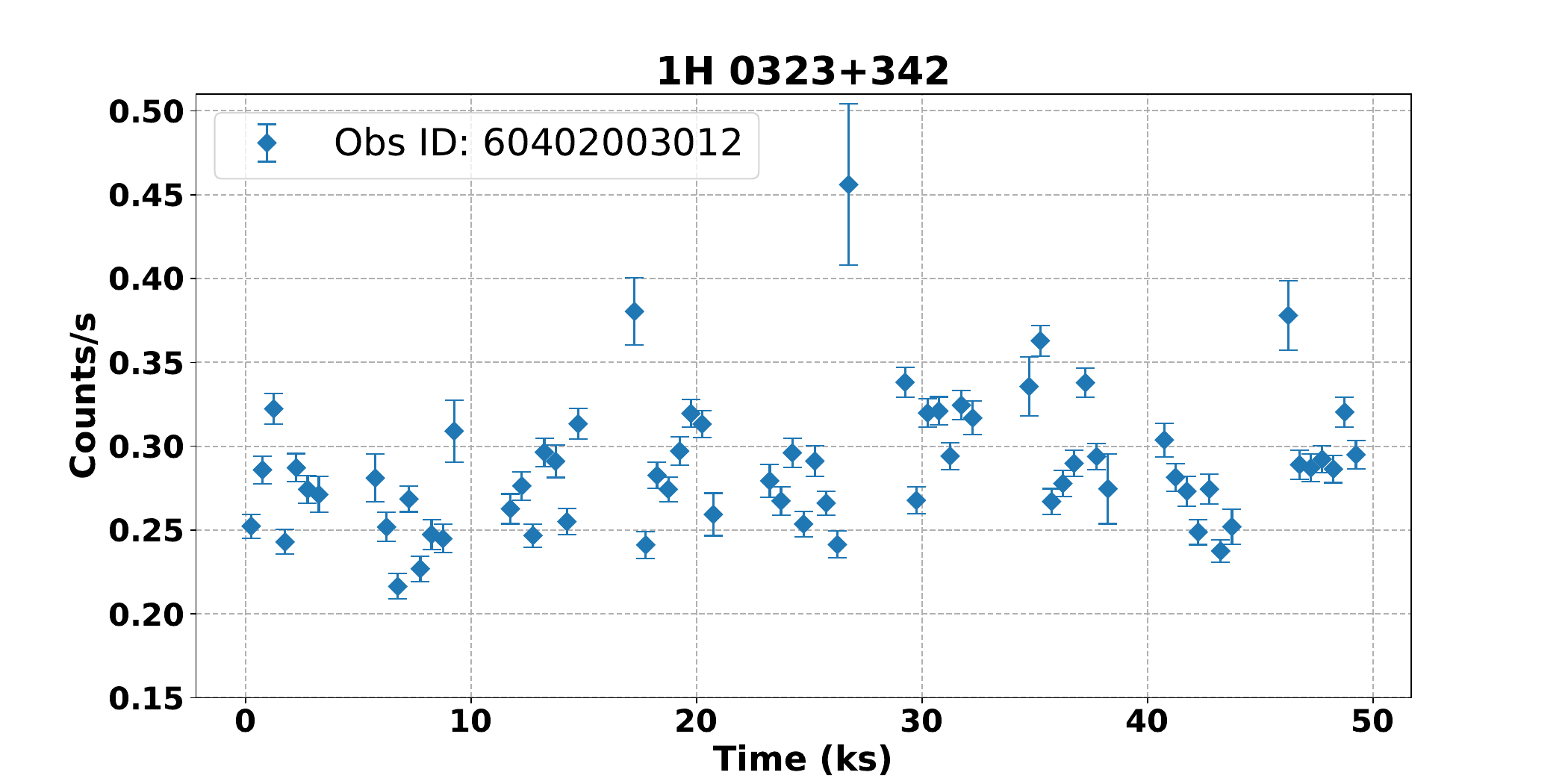}
       \includegraphics[angle=0, width=0.480\linewidth]{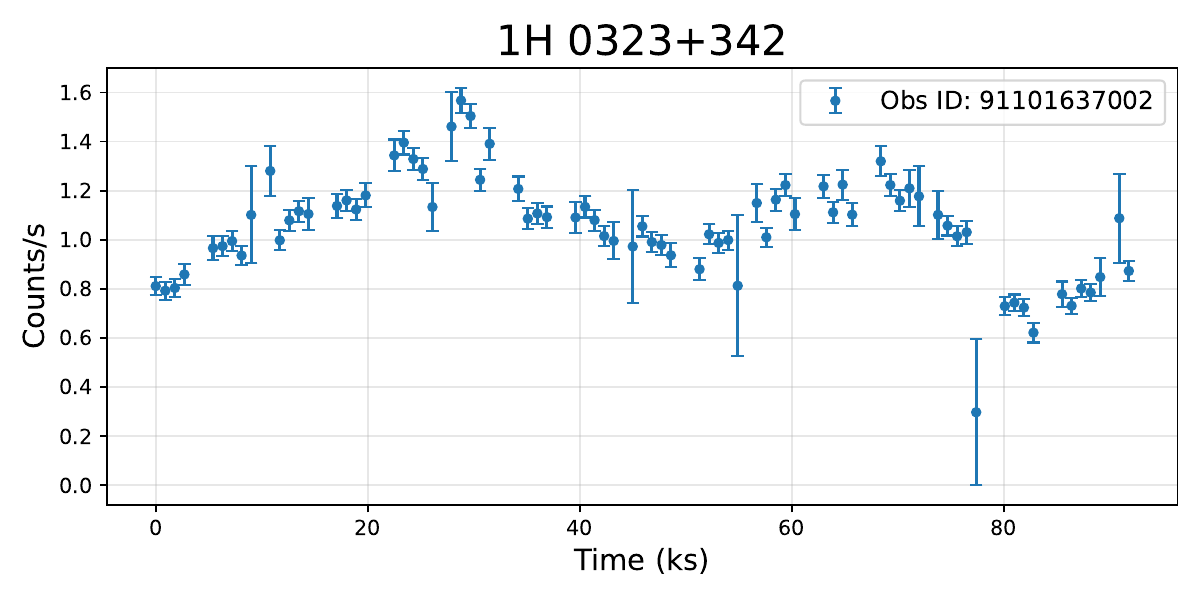}
    \caption{X-ray (NuSTAR) lightcurves of 1H 0323+342.}
    \label{LCs}
\end{figure*}

The computed $F_{var}$, listed in Column 4 of Table \ref{tab: NuSATAR_info} and calculated through Equation \ref{FVs}, reveals moderate variability of the light curves. Throughout our sample, the mean $F_{var}$ is around 13$\%$, with the highest being 22$\%$ for 1H 0323+342 and the lowest at 10$\%$ for PMN J0948+0022. Based on Swift observations in \cite{2020MNRAS.496.2213D}, the broadband timing analysis of a sample of $\gamma$-NLSy1 galaxies reveals a moderate level of $F_{var}$. In the X-ray band $F_{var}$, varies from $ 0.27 \pm 0.07$ for RGB J1644+2619 to $0.51 \pm 0.08$ for PKS 1502+036. While the $F_{var}$ in the optical band shows a greater extent of flux variations, from 0.10$\pm$0.01 for 1H 0323+342 to 1.44$\pm$0.04 for SBS 0846+513. Similarly, in the UV band, 1H 0323+342 has the minimum $F_{var}$ at 0.15$\pm$0.01, whereas SBS 0846+513 has the maximum variability at 1.06$\pm$0.04. The hard X-ray light curves of $\gamma$-NLSy1 galaxies are presented in Figure \ref{LCs}.

For the NLSy1, F$_{var}$ ranges between 10-20\% as can be seen in Table 2, which is very narrow in range and low in value compared to FSRQs and BL Lacs. \cite{2018A&A...619A..93B} have studied the hard X-ray sample of FSRQs and BL Lacs, and they have shown that the estimated F$_{var}$ for FSRQs lies in the range between 5-30\% roughly, and in the case of BL Lacs it is between 5-40\% suggesting FSRQs and BL Lacs inherently have more variability than NLSy1 galaxies.

\subsection{OPTICAL BRIGHTNESS \& COLOUR VARIABILITY}
We searched for the optical long-term data of our sample sources from the recent Zwicky Transient Facility (ZTF) public data release\footnote{\url{https://www.ztf.caltech.edu/ztf-public-releases.html}} (see, \cite{2019PASP..131a8003M} and found five sources in the ZTF survey (except PKS 2004-447), within a positional uncertainty of 1.5$^{''}$. We have presented the extinction-corrected g and r-band long-term lightcurves in Figure \ref{ZTFLCs} during MJD 58000 to 61000.

To quantify the ZTF optical variability in g- and r-band lightcurves, we have used fractional variability as discussed in the above Equation \ref{FVs} and amplitude of variability, $\psi$ discussed in \cite{1996A&A...305...42H}:
\begin{equation}
    \psi = \sqrt{((A_{max} - A_{min})^2 - 2\sigma^2},
\end{equation}
where, $A_{max}$ and $A_{max}$ are maximum, and minimum amplitude in the lightcurves and $\sigma^2 = <\sigma_{i}^2>$, $\sigma_i$ is the error in the $i^{th}$ data point. Our analysis reveals that both the r- and g-band lightcurves show a significant variability for all the sources presented in Table \ref{TAB3}. The amplitude of variability, $\psi$, ranges from 1.11 to 2.31, with a mean value of 1.66 in r-band and 0.90 to 2.32, with a mean of 1.67, and $F_{var}$ varies from 13 to 40\% in r-band, while in g-band it ranges from 11 to 44\%. A systematic study of blazars using ZTF lightcurves presented in  \cite{2022MNRAS.510.1791N} reveals that the BL Lacs are more variable than FSRQs. The observed $F_{var}$ in optical bands is much higher than in the X-ray band, suggesting the optical photons originate within the inner jets and disc. 

It is obvious that in the blazar jet emission dominates the disk in optical because of the presence of strong jets, and they are entangled together, so difficult to separate those emissions. However, on the other hand, as seen from the broadband SED in NLSy1, the disk can have a significant contribution, and in principle, it can be disentangled from the jet. The color-magnitude variability can be used to distinguish the jet and the disk emission. In \cite{Bonning_2012}, they show that most of the FSRQs show a RWB trend since the emission is highly dominated by redder jet emission. On the other hand, \cite{10.1093/pasj/63.3.327} argues that the bluer-when-brighter is more common in BL-Lac type of objects.
Along with the optical variability, we have also plotted the color-magnitude diagram of our sample sources. For this, we have collected quasi-simultaneous g- and r-band data within an hour. The color-magnitude plots are presented in Figure \ref{COLORS}. Our analysis suggests a RWB trend for most of our sample, which is also a well-known feature of blazar classes FSRQ. This finding suggests that gamma-ray-detected NLSy1 galaxies are much closer to FSRQs than BL Lacs. Our analysis shows that $\gamma$-NLSy1 galaxies exhibit a mix of disk-related and jet-driven variability in their optical variability. Some (like PKS 1502+036) have strong, consistent variability akin to blazars, while others (like J1222+0413) exhibit more stochastic behavior that may be influenced by an accretion disk. The disc contribution of NLSy1s is higher than that of BL Lac objects, which exhibit stronger and faster variability due to pure jet domination. Although NLSy1s sometimes exhibit increased g-band variability, reflecting their low black hole masses and high accretion rates \citep{2019ApJ...881L..24V}, FSRQs exhibit higher variability amplitudes and redder-band dominance due to their enormous black holes and powerful jets. This establishes a connection between thermal disc and non-thermal jet activity, making $\gamma$-NLSy1s a category in between blazars and Seyfert galaxies.

\begin{table*}
    \centering
    \begin{tabular}{c|cc|cc|c}\hline
        Name & $\psi_r$ & $F_{var,r}$ & $\psi_g$ & $F_{var,g}$ & $\Delta F_{var}$ \\ \hline
        1H 0323+342 & 1.3584 ± 0.0045 & 0.1362 ± 0.0002 & 0.8950 ± 0.0075 & 0.1598 ± 0.0004  & 0.0236 \\
       PKS 1502+036  & 2.3074 ± 0.0160 & 0.4035 ± 0.0013  & 2.3160 ± 0.0238 & 0.4366 ± 0.0024  & 0.0331 \\
       J1644+263 & 1.5623 ± 0.0072 & 0.2123 ± 0.0005  & 1.6293 ± 0.0079 & 0.2031 ± 0.0007  & -0.0092 \\
        J0948+0022 & 1.9357 ± 0.0158 & 0.3153 ± 0.0017 & 1.8494 ± 0.0165 & 0.2259 ± 0.0023 & -0.0894 \\
       J1222+0413  & 1.1156 ± 0.0086 & 0.1493 ± 0.0013  & 1.1649 ± 0.0078 & 0.1119 ± 0.0017 & -0.0374 \\ \hline
    \end{tabular}
    \caption{ZTF light curve variability in r- and g-band}
    \label{TAB3}
\end{table*}

\subsection{SPECTRAL ANALYSIS}
Here, we have performed the hard X-ray spectral analysis by fitting a power-law (PL) in the energy range 3.0-79.0keV in \emph{XSPEC}. The PL model is described as:
\begin{equation}\label{EQ1}
    \frac{dN}{dE} = N\cdot E^{-\Gamma},
\end{equation} 
where $N$ and $\Gamma$ represent the normalization constant and hard X-ray photon index, respectively. Our observations are well-fitted by the PL model, suggesting a non-thermal hard X-ray emission originating from the jet and corona. The results of the X-ray spectral fitting are summarized in Table \ref{tab: NuSATAR_info}. The analysis indicates that 1H 0323+342 is the most luminous source in our sample, with an average unabsorbed flux of (26.66$\pm$0.35)$\times 10^{-12}$ erg cm $^{-2}$ s$^{-1}$. However, across eight NuSTAR observational epochs, its flux varies significantly, ranging from a minimum of 14.08$\pm$0.24 (for observation ID 60402003012) to a maximum of 72.07$\pm$0.70 (for observation ID 91101637002) $\times 10^{-12}$ ergcm$^{-2}$s$^{-1}$. The X-ray photon index remains nearly constant across all epochs, averaging $\Gamma = 1.81$, which suggests a hard spectral nature. PKS 2004-447 has been observed three times with NuSTAR. Its spectral index averages $\Gamma = 1.55$, while its flux varies between a minimum of 3.09$\pm$0.12, an average of 4.13$\pm$0.18, and a maximum of 5.73$\pm$0.30. PKS 1502+036, RGB J1644+263, PMN J0948+0022, and CGRaBS J1222+0413 have only single NuSTAR observations. Among all the sources in our sample, PKS 1502+036 exhibits the lowest flux 1.98$\pm$0.14, and the hardest spectrum with $\Gamma = 1.23$. A soft excess and Fe$K_{\alpha}$ line were consistently found at lower energies in previous hard X-ray spectral fitting analyses \citep{2018MNRAS.479.2464G, 2020MNRAS.496.2922M} that used data from XMM-Newton, Swift, Suzaku, and NuSTAR. At higher energies, there were signs of a hard excess. This hard excess was perceived as a component of blurred reflection. In particular, \cite{2018MNRAS.479.2464G} used source and background regions of 60 and 80 arcseconds, respectively, and required a minimum of 100 counts per energy bin and noted a hard excess in 1H 0323+342 above $\sim$35 keV. On top of that, \cite{2020MNRAS.496.2922M} chose smaller source regions of 40 arcseconds and at least 25 counts per energy bin, reporting a hard excess above $\sim$40 keV. However, there is no overt indication of a hard excess in our independent study, which was carried out using the same source and background regions and binning criteria from these earlier investigations. Rather, we discover that a straightforward power-law model may adequately represent the NuSTAR spectra without the need for an extra reflection component, suggesting non-thermal jet emission. To ascertain whether these observable discrepancies are the result of methodological decisions or reflect actual astrophysical variability, more research is required.
\cite{2019A&A...632A.120B} used Swift-XRT, XMM-Newton, and NuSTAR data to present a comprehensive X-ray investigation of four $\gamma$-NLSy1 galaxies and concluded that nonthermal emission from the relativistic jet dominates the spectra above $\sim$2 keV in all four sources. The resulting photon indices are much harsher than those seen in radio-quiet NLS1s and are similar to those commonly found in FSRQs. Furthermore, a soft X-ray excess below 2 keV is seen in three of the four sources, which is probably related to thermal Comptonization in a hot corona. The broadband spectrum for 1H 0323+342 was well represented by a broken power-law model with a break energy of $E_{\rm break} \sim 13.4$ keV and photon indices of $\Gamma_1 = 1.83 \pm 0.02$ and $\Gamma_2 = 1.68 \pm 0.07$. PKS 2004-447 and J0948+0022, on the other hand, were better fitted with a {\textit{PL+compTT}} model, resulting in parameters of $kT = 21.5^{+54.2}_{-16.1}$ keV, $\Gamma = 1.60 \pm 0.04$ and $kT \geq 2.1$ keV, $\Gamma = 1.31 \pm 0.04$, respectively. However, a single power-law proved sufficient to characterize PKS 1502+036, with $\Gamma=1.16 \pm 0.08$. Although we did not include the soft X-ray band in this work, this precludes a direct comparison with our results.

\section{RESULTS AND DISCUSSION} \label{discussion}
AGNs are classified based on their viewing angle and optical spectral properties. In this work, we investigate the temporal and spectral X-ray properties of selected gamma-ray detected NLSy1 galaxies and try to answer fundamental questions such as how the NLSy1 galaxies are different from other AGN types, such as FSRQs and BL Lacs. We collected the archival NuSTAR observations of six gamma-ray-detected NLSy1 and produced the light curve and spectra. The light curve reveals the short-term variability present in the sources, and the variability is quantified based on the estimation of fractional variability amplitude (F$_{var}$). 
For the NLSy1, F$_{var}$ estimated for X-rays ranges between 10-20\% as can be seen in Table \ref{tab: NuSATAR_info}, which is very narrow in range and low in value compared to FSRQs and BL Lacs. \cite{2018A&A...619A..93B} have studied the hard X-ray sample of FSRQs and BL Lacs, and they have shown that the estimated F$_{var}$ for FSRQs lies in the range between 5-30\% roughly, and in the case of BL Lacs it is between 5-40\% suggesting FSRQs and BL Lacs inherently have more variability than NLSy1 galaxies. This suggests that even though the X-rays are produced in the jet, they are less variable than FSRQs and BL Lacs. This can be due to the difference between the kinetic energy of the electrons present in the jets or due to the lower value of the magnetic field in NLSy1 (considering that soft X-rays are mostly produced by the synchrotron-self Compton mechanism). This has also been seen in \cite{2019ApJ...872..169P}, where the peak of the particle energy distribution ($\gamma_b$) for some of the NLSy1 is derived close to 2000, whereas, for blazars, it goes beyond 10,000. 

The X-ray spectral analysis reveals that the PL index for 1H 0323+342 is consistent throughout the flux state, mostly close to 1.81, whereas for PKS 2004-447 is between 1.36 to 1.66. For PKS 1502+036, the photon index is 1.23, and for RGB J1644+263 is 1.75. For PMN J0948+0022 and CGRaBS J1222+0413 is 1.45. In all the spectra, we noticed that a single power-law is sufficient enough to fit the X-ray spectra from 3-50 keV (in most cases) without any Comptonization component, suggesting the X-rays are produced in the jet similar to FSRQs and BL Lacs.

\begin{figure*}
\includegraphics[height=0.265\linewidth,width=5.5cm]{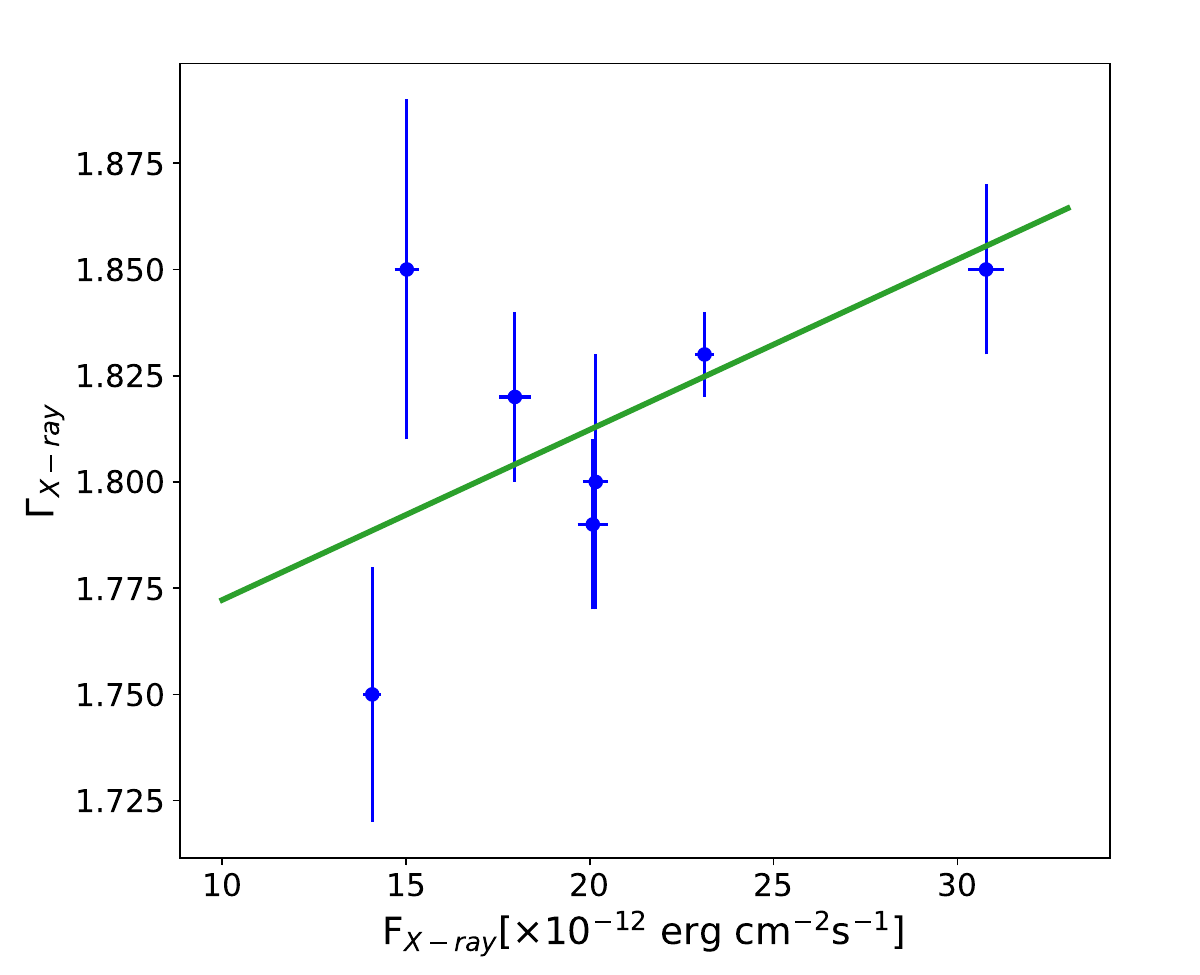}
  \includegraphics[height=0.25\linewidth,width=5.8cm]{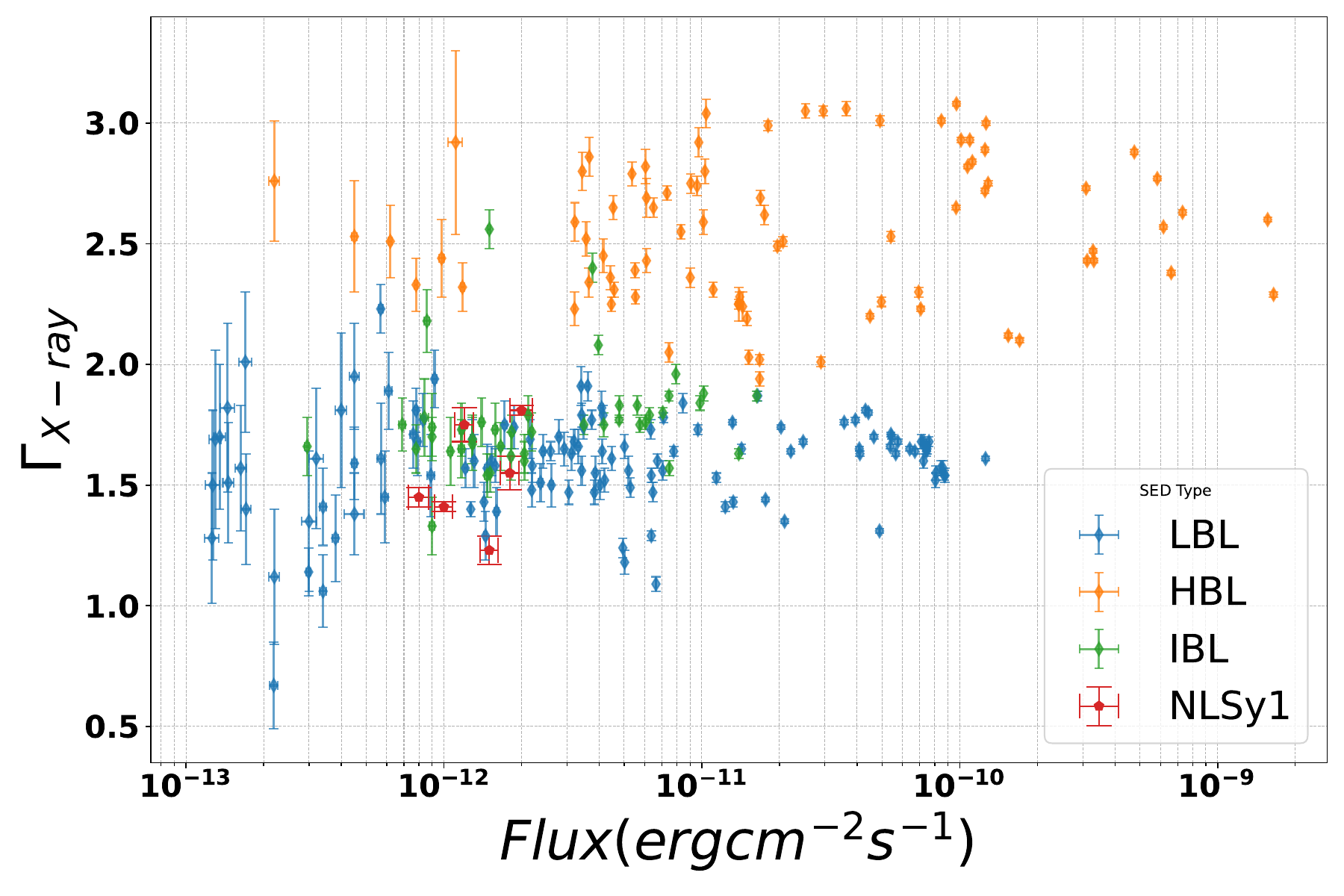}
     \includegraphics[height=0.252\linewidth,width=6.5cm]{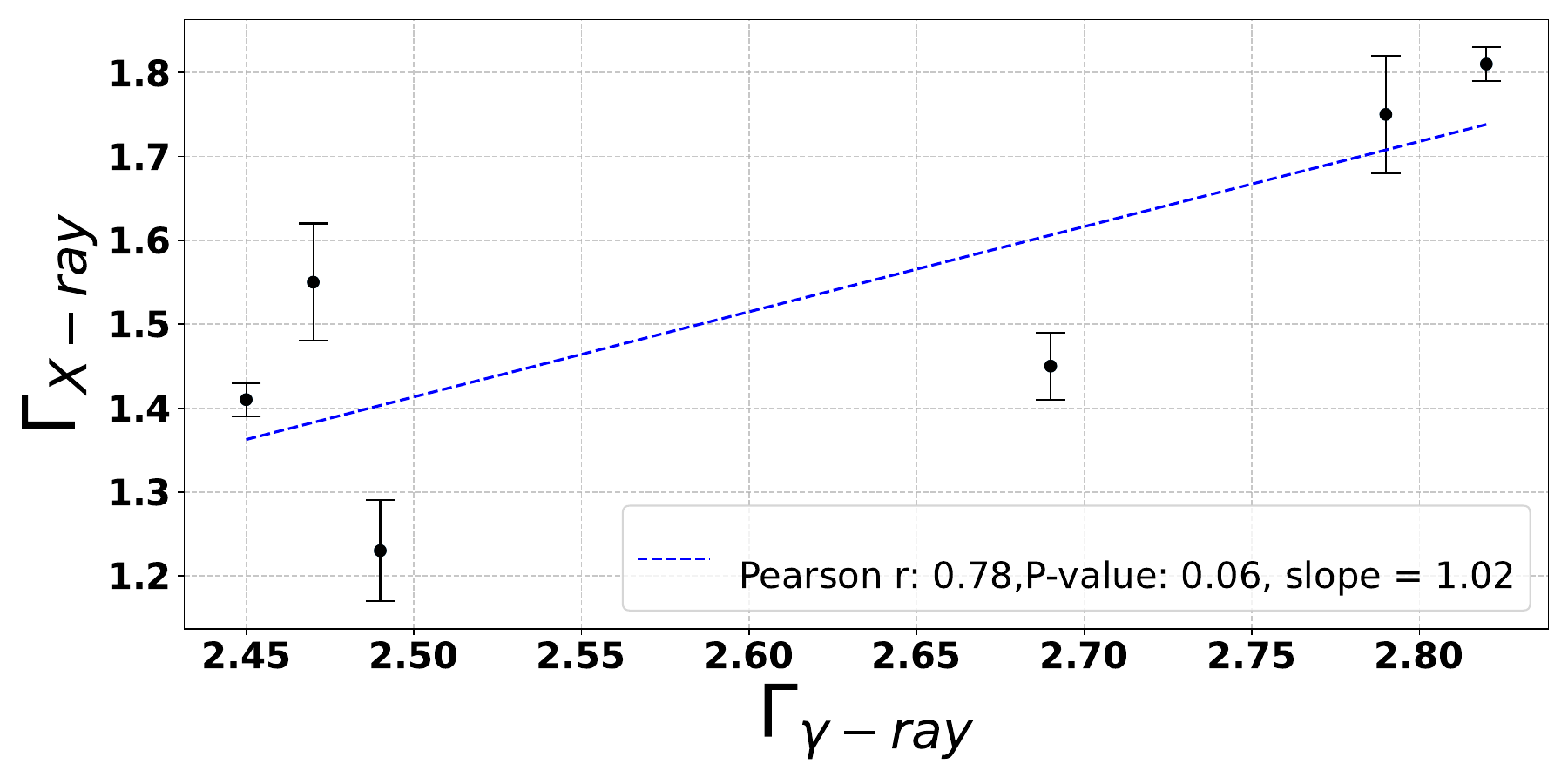}
\caption{The flux vs index plot is shown for 1H 0323+342. It shows the softer-when-brighter behavior similar to a blazar \cite{2018A&A...619A..93B}. Comparison of Hard X-ray Photon Index vs. Hard X-ray Flux for NLSy1 Galaxies and Blazar Subclasses (LBL, IBL, HBL). Hard X-ray Photon Index vs. Average Gamma-ray Photon Index from Fermi 4FGL.}
\label{c1}
\end{figure*}

The SED modeling performed in \cite{2019ApJ...872..169P} shows that the X-ray emission lying in the external Compton peak is most probably explained by the inverse-Compton scattering of BLR photons. In Figure \ref{c1}, we show the X-ray spectral index variation for 1H 0323+342 with X-ray flux, and we noticed a softer-when-brighter trend similar to FSRQs. Based on the broadband SED and the X-ray photon spectral index, we can conclude that the NLSy1 galaxy 1H 0323+342 is closer to FSRQ and Low BL Lacs objects. 

We also plot the X-ray spectral index of all the sources along with BL Lac samples, and it shows that NLSy1 are more like low BL Lac objects (middle panel of Figure \ref{c1}). In the other plot shown in Figure \ref{c1}, we compare the X-ray and gamma-ray spectral index for our sample, and a moderate positive correlation is observed, suggesting X-ray and gamma-rays are produced by the same population of electrons and through the same processes.  

\begin{figure*}
    \centering
    \includegraphics[width=0.48\linewidth]{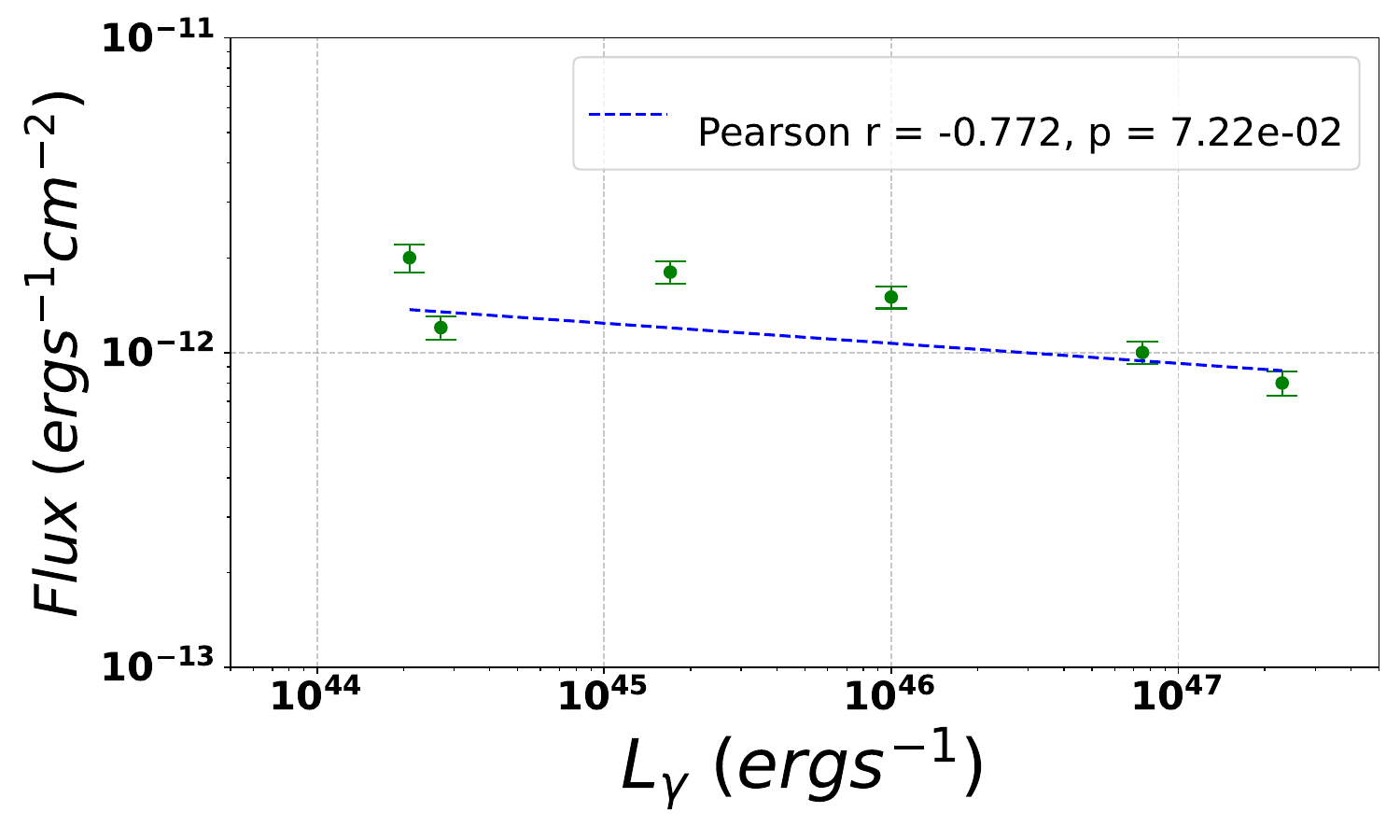}
    \caption{Comparison of Hard X-ray flux vs $L_{\gamma}$. }
    \label{c2}
\end{figure*}

\begin{figure*}
    \centering
    \includegraphics[width=0.48\linewidth]{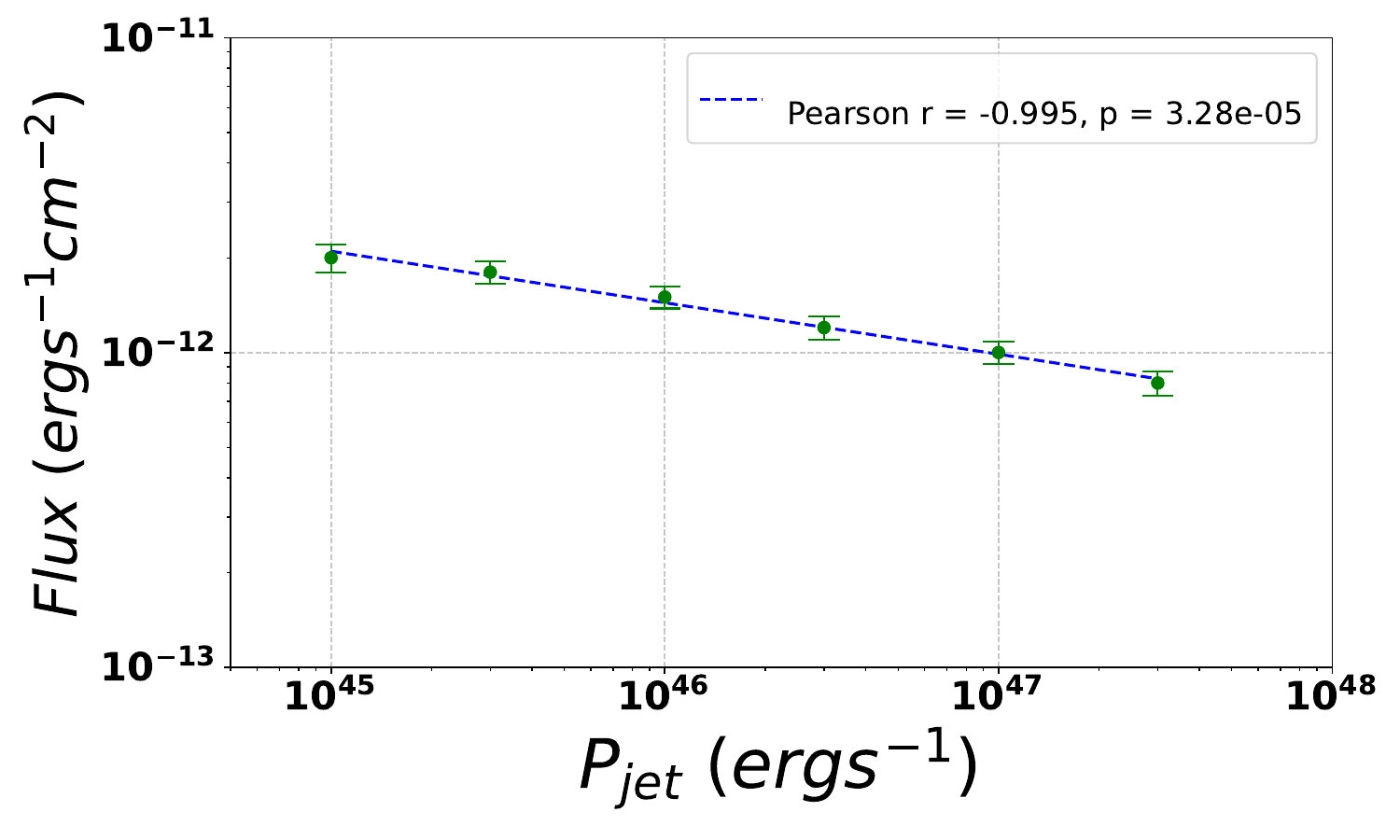}
    \includegraphics[width=0.48\linewidth, height=5.6cm]{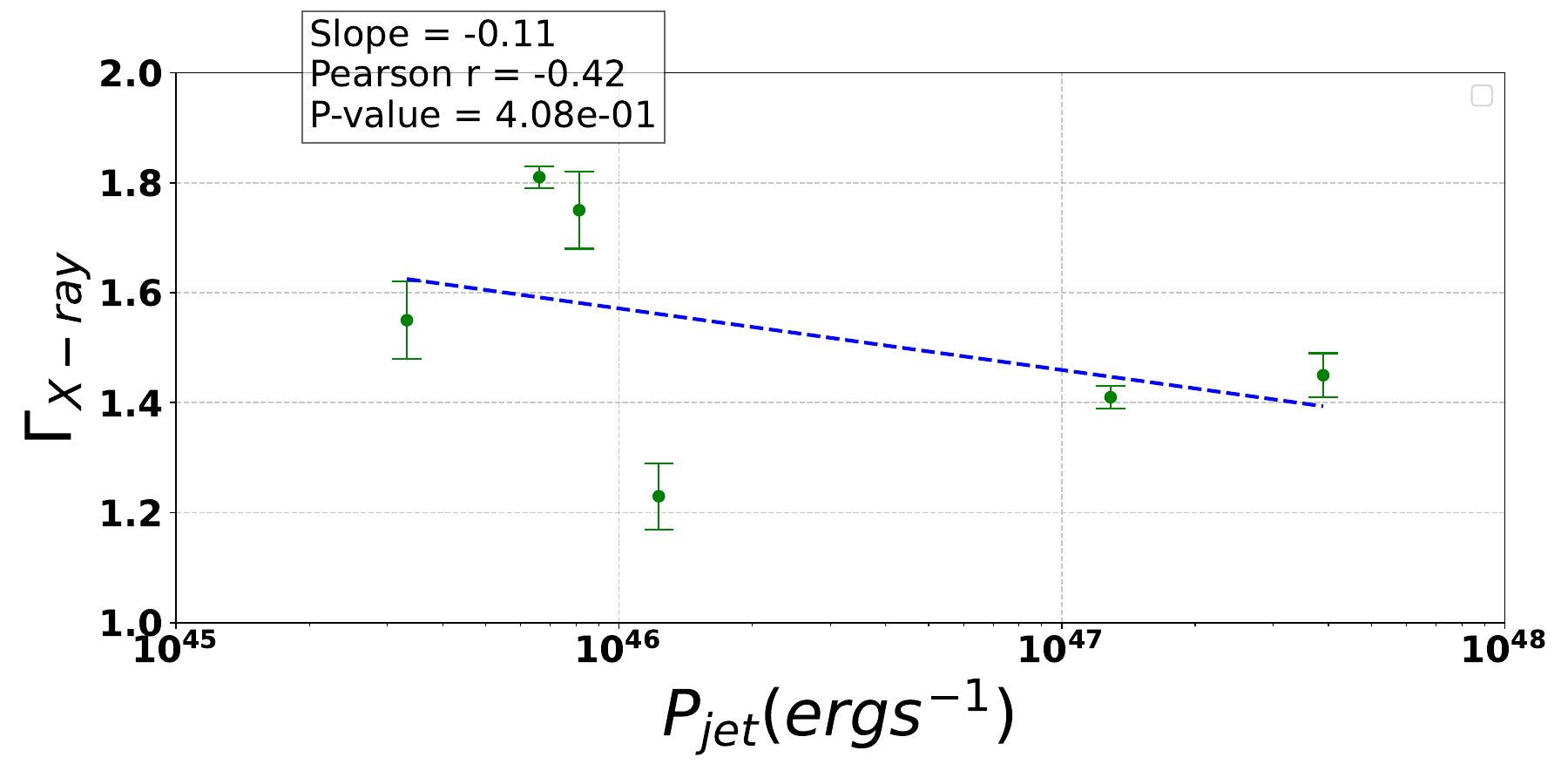}
    \caption{Correlation Between Hard X-ray Flux, X-ray Photon Index, with Jet Power.}
    \label{c3}
\end{figure*}

In Figure \ref{c2}, we plot the gamma-ray luminosity vs the X-ray flux, and a moderate anti-correlation is seen for our sample, suggesting gamma-ray flares are somewhat stronger than the X-ray flares. Combining this information with Figure \ref{c3}, where the X-ray flux shows an anti-correlation with total jet power, we conclude that the jet power is more dominated by the gamma-ray radiation. A weak anti-correlation of the X-ray photon spectral index is seen with total jet power. 

\begin{figure*}
    \centering
    \includegraphics[width=0.48\linewidth, height=5.8cm]{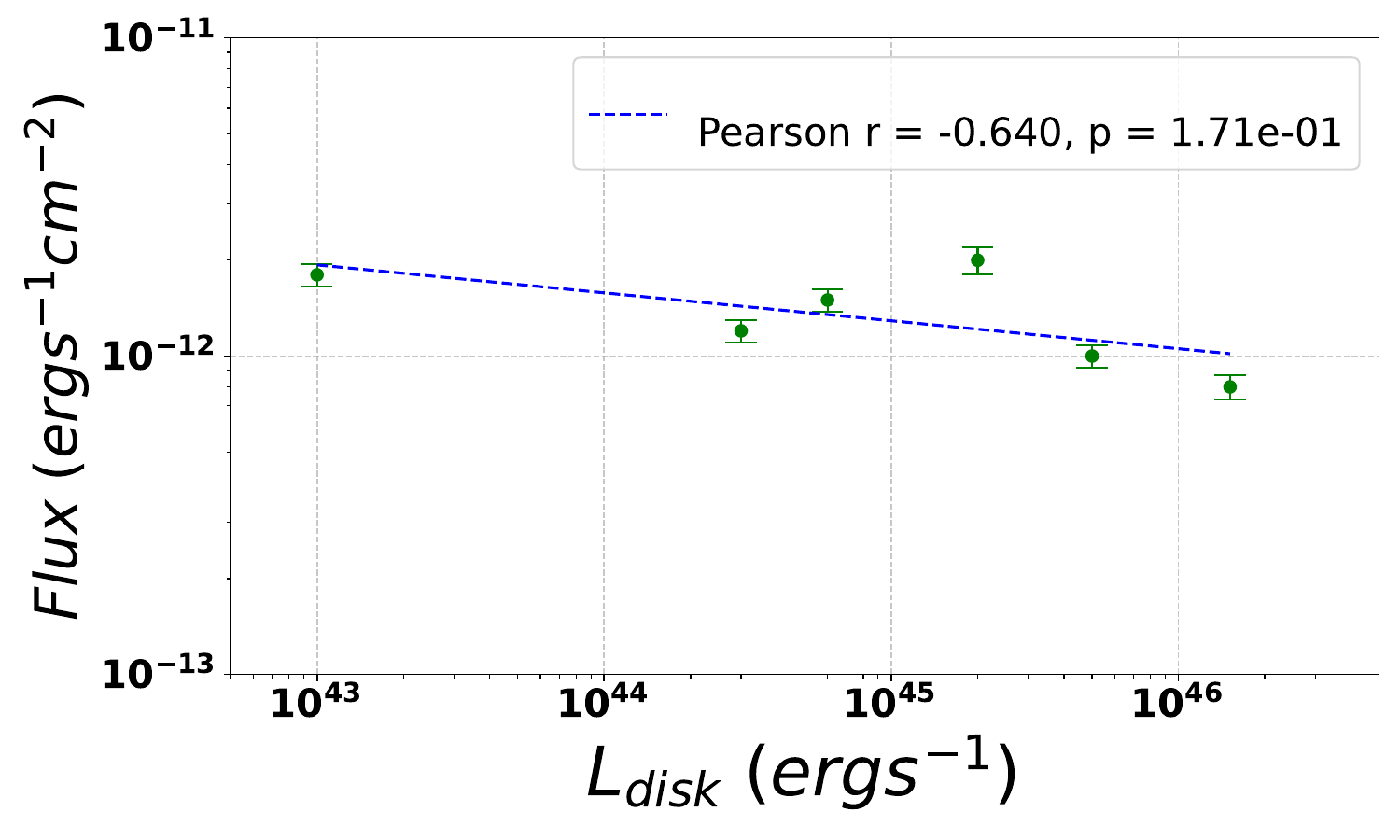}
    \includegraphics[width=0.48\linewidth]{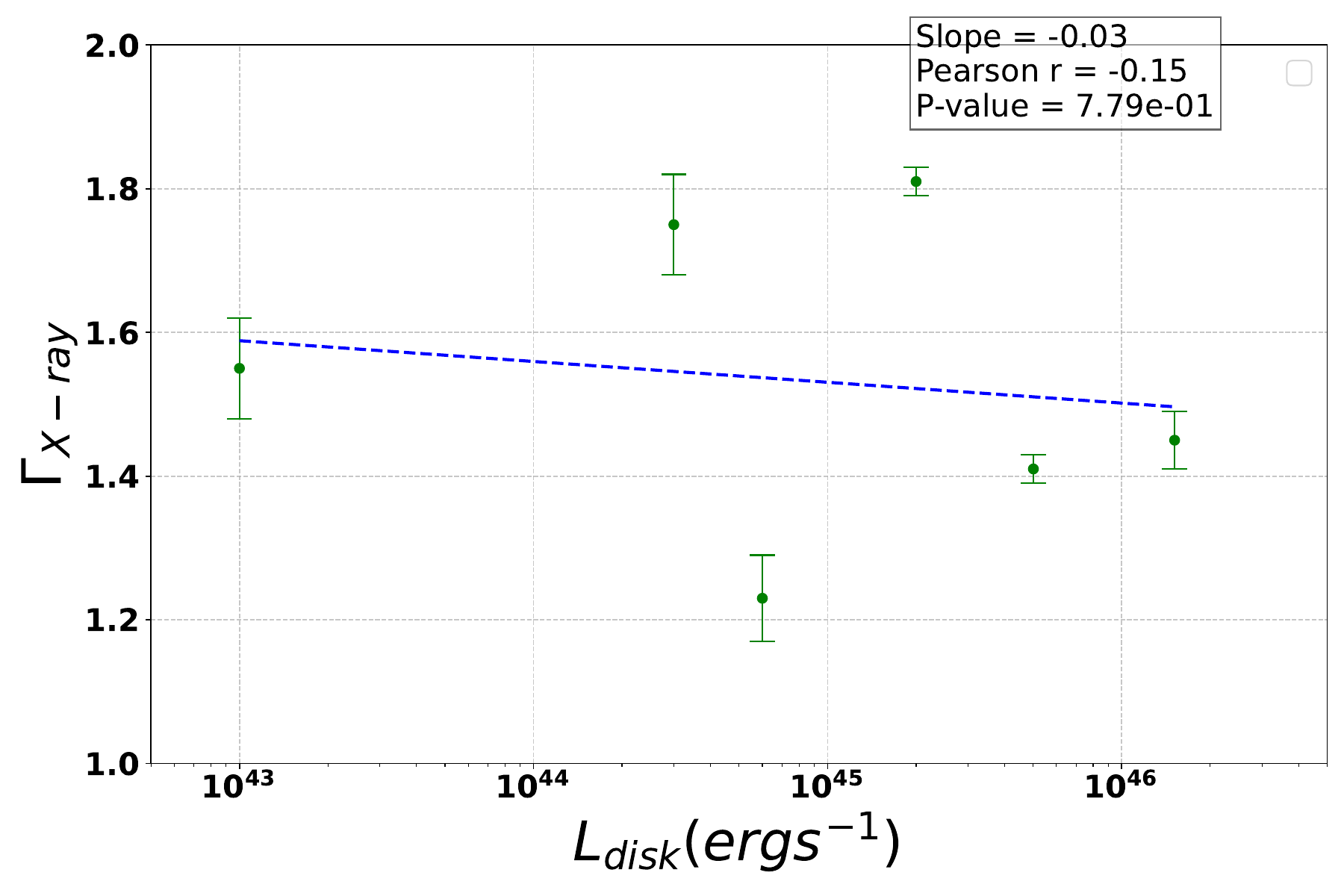}
    \caption{Correlation Between Hard X-ray Flux, X-ray Photon Index, with Disk Luminosity.}
    \label{c4}
\end{figure*}

In Figure \ref{c4}, we show the X-ray flux vs disk luminosity plot, and a moderate anti-correlation is observed, suggesting that the object with stronger disk emission can suppress the jet emission since the NLSy1 has a larger viewing angle and the disk is easily visible, unlike blazars. The X-ray photon spectral index does not show any significant correlation with the disk luminosity. An important point to note here is that the X-ray flux and index used in these figures are non-simultaneous with the parameter values used here, such as jet power or disk luminosity. These might affect the correlation. This is also true for Figure \ref{c1} and \ref{c2}, where we correlate the X-ray and gamma-ray properties.

%The average gamma-ray spectral index for 1H 0323+342 is 2.818 (4FGL-DR4), and the average hard X-ray spectral index is 1.813. 

\begin{figure*}
    \centering
    \includegraphics[height=0.36\linewidth,width=8.0cm]{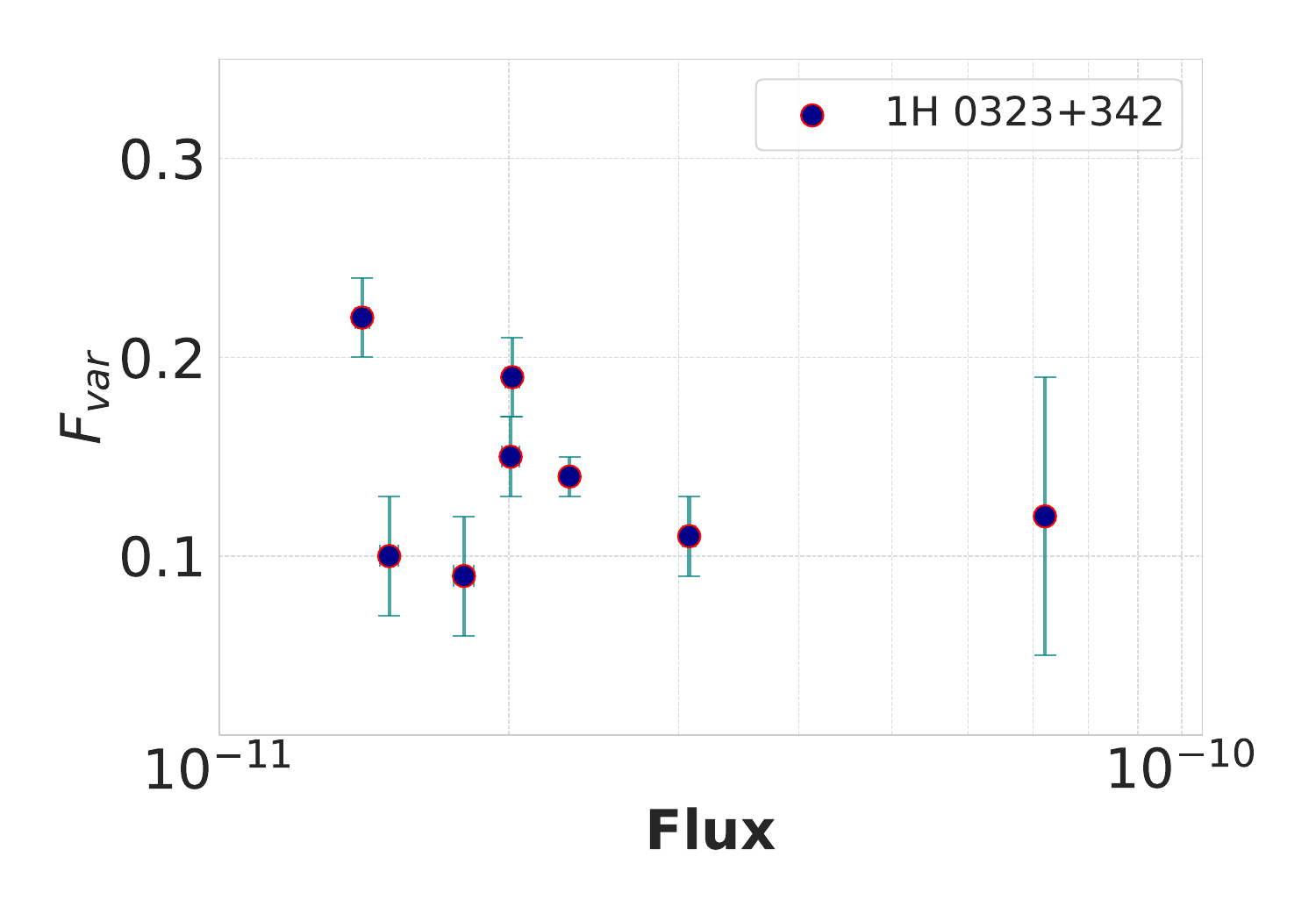}
    \includegraphics[height=0.35\linewidth,width=8.0cm]{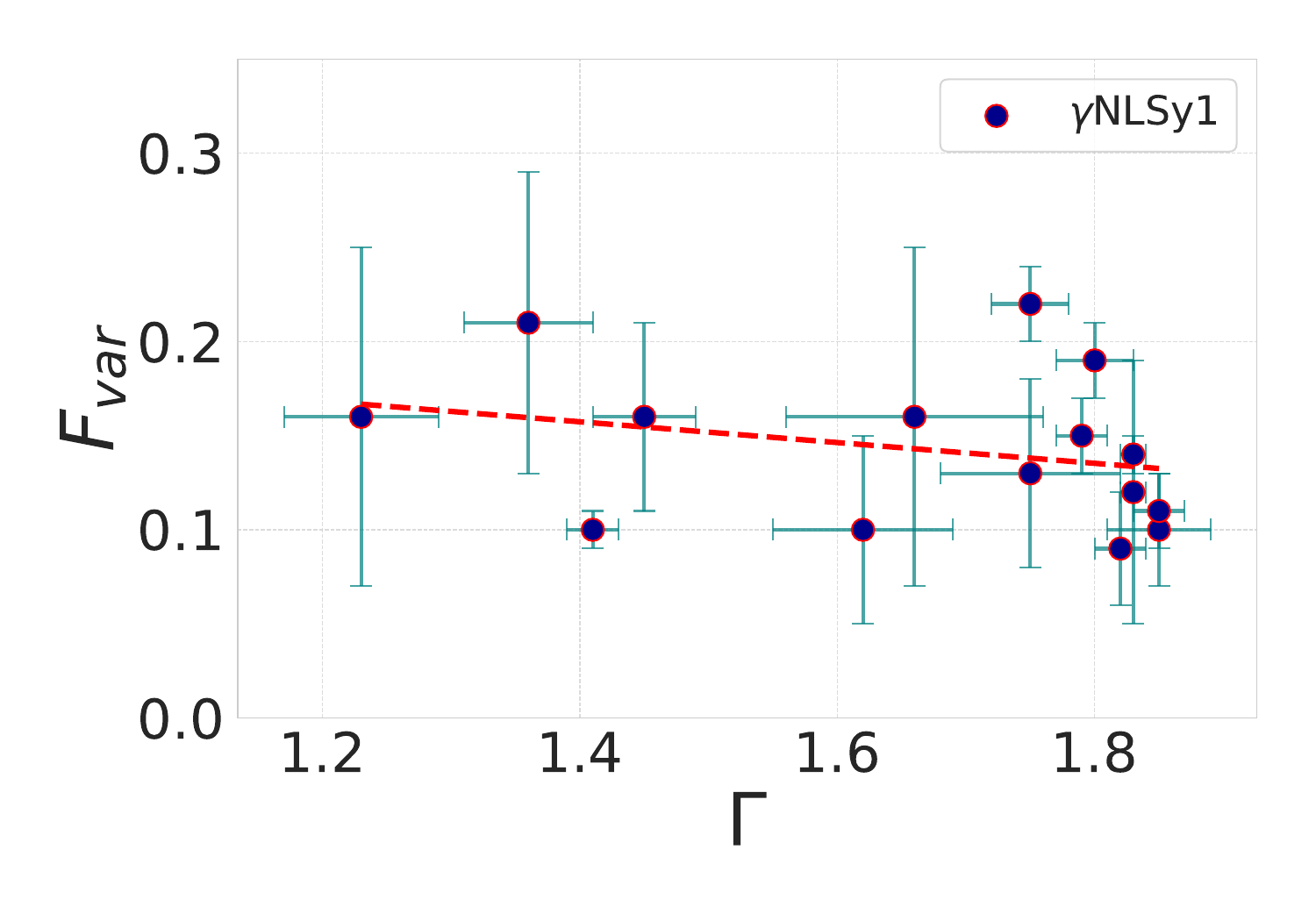}
    \caption{Left panel: Hard X-ray Flux (erg cm$^{-2}$ s$^{-1}$) vs $F_{var}$ for 1H 0323+342; Right panel: $F_{var}$ vs photon index of all sources.}
    \label{c5}
\end{figure*}

We have also checked if there is any trend that exists between the X-ray photon index and the F$_{var}$ for our sample of NLSy1, but we do not observe any trend. However, we plot the F$_{var}$ concerning flux for 1H 0323+342, and we observed an anti-correlation. In Figure \ref{c5}, we also show the F$_{var}$ variation with the hard X-ray spectral index. A mild hint of anti-correlation is seen. Comparing the F$_{var}$ vs $\Gamma$ in \cite{2018A&A...619A..93B} we observed that both FSRQs and BL Lacs show a clear positive correlation and the trend is clearer in the case of FSRQs whereas, in case of NLSy1, the trend is mild and opposite suggesting that as the flux increases the spectrum becomes steeper revealing faster cooling of electrons.

\begin{figure*}
    \centering
    \includegraphics[width=0.99\linewidth]{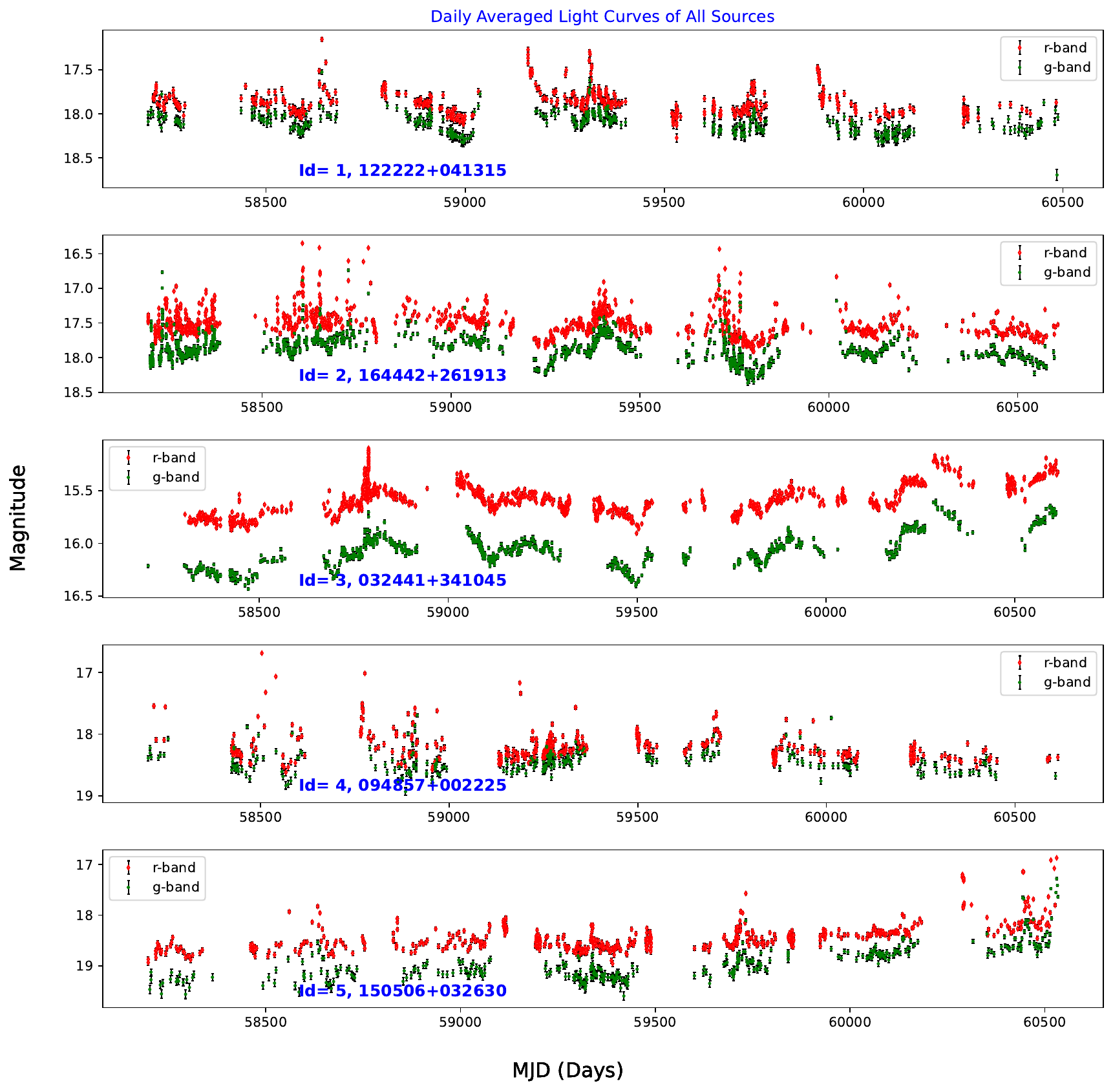}
       \caption{ZTF r- and g-band lightcurves of our sample sources}
       \label{ZTFLCs}
\end{figure*}

\begin{figure*}
    \centering
    \includegraphics[width=0.45\linewidth]{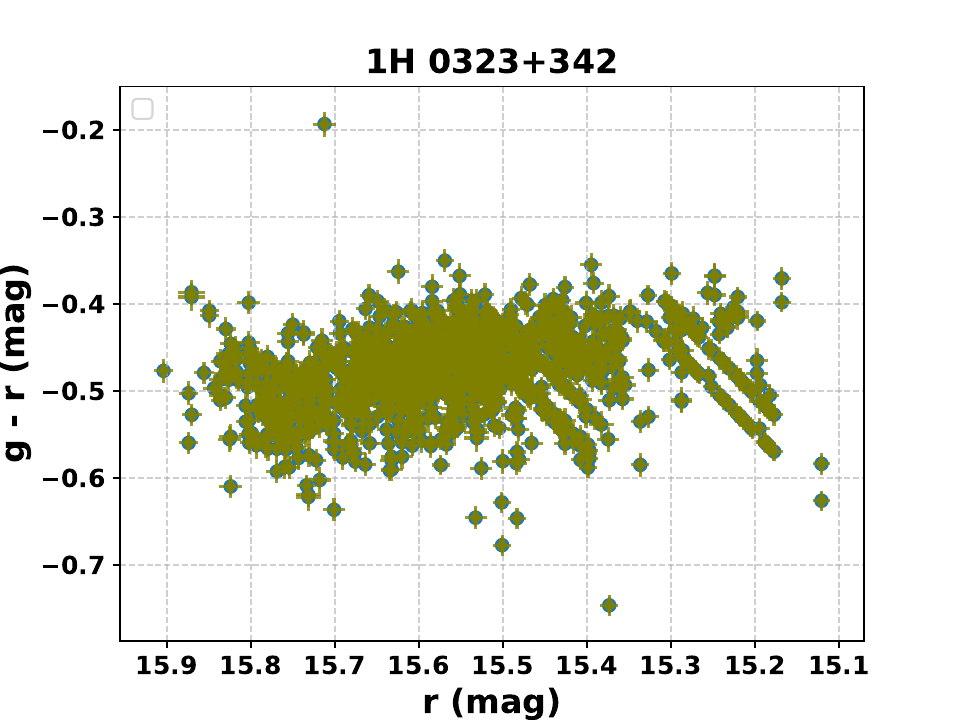}
    \includegraphics[width=0.45\linewidth]{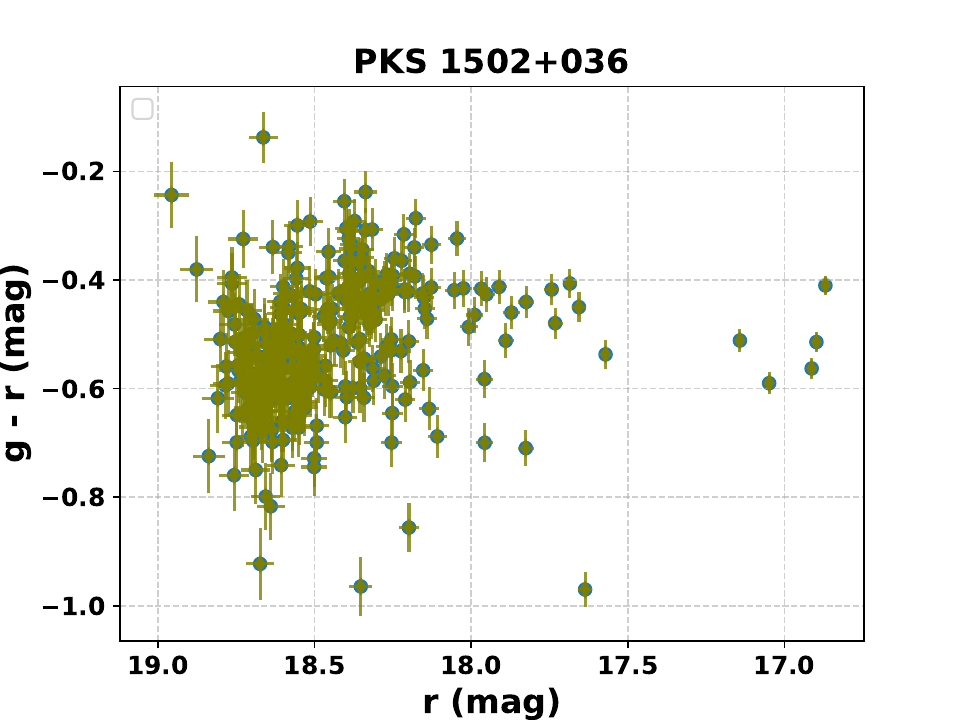}
    \includegraphics[width=0.45\linewidth]{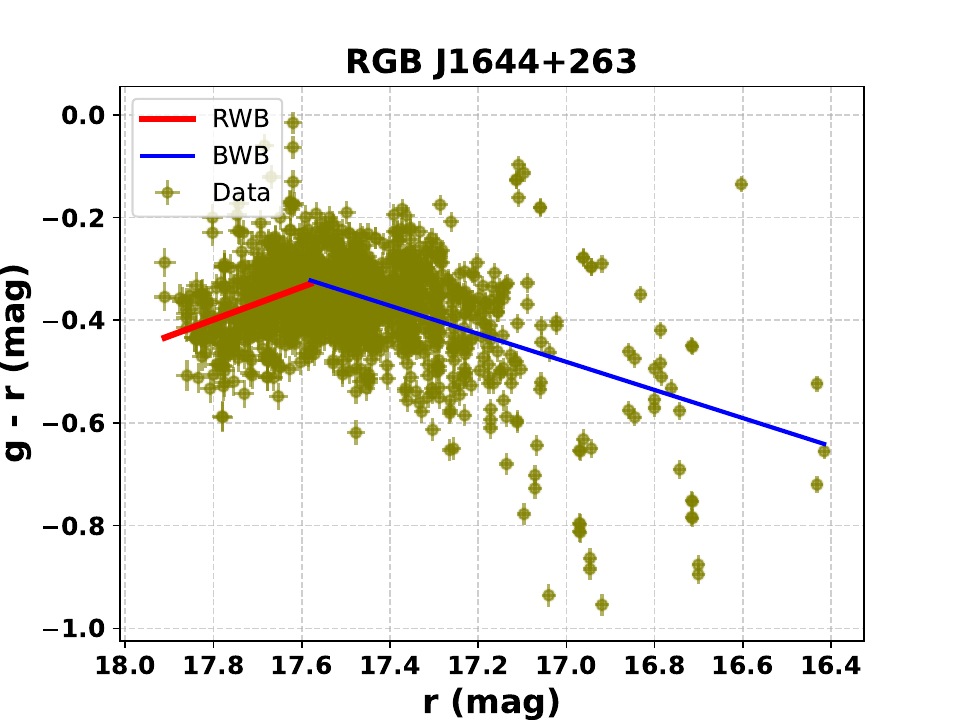}
    \includegraphics[width=0.45\linewidth]{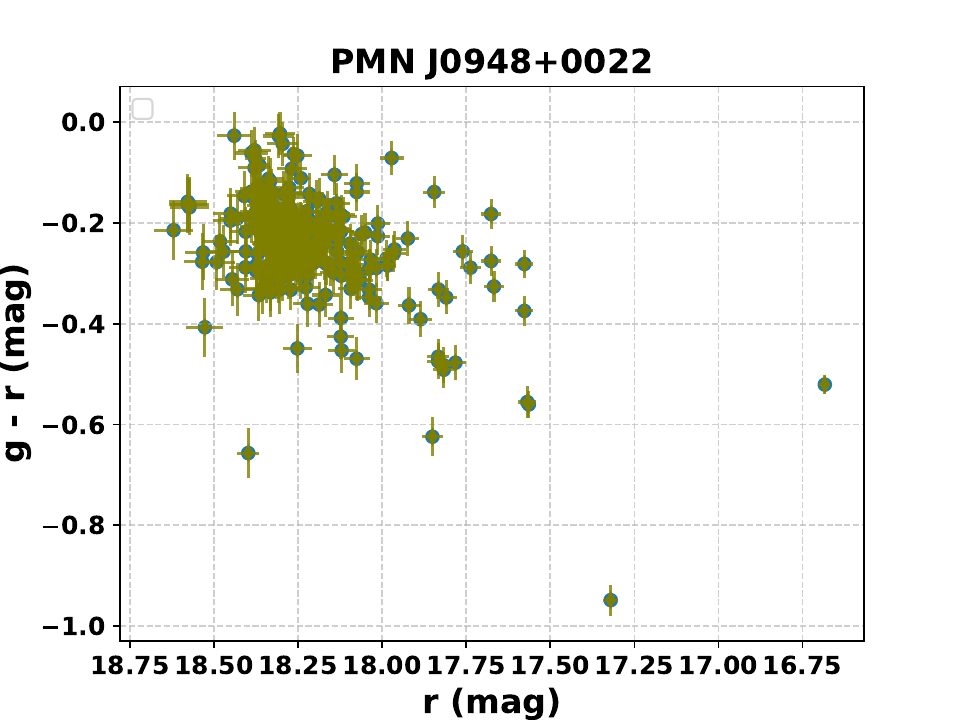}
    \includegraphics[width=0.45\linewidth]{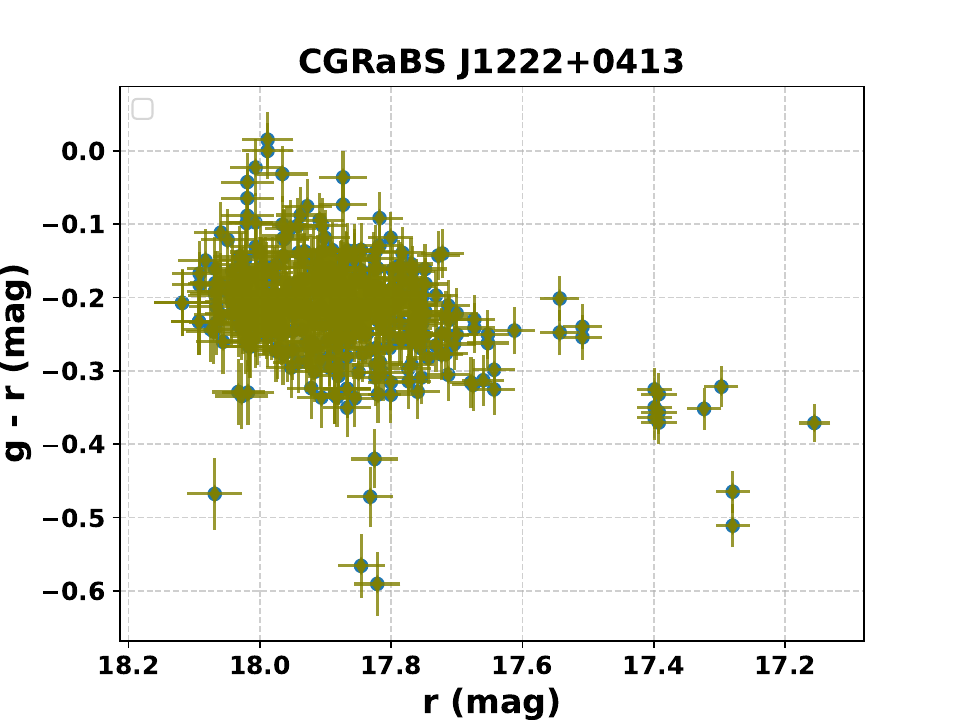}
    \caption{Our sample's optical color-magnitude diagram (g-r vs. r) using near-simultaneous photometric data within 1 hour.}
    \label{COLORS}
\end{figure*}

We collected the g- and r-band ZTF light curves for five objects, which are available in the ZTF archive, and showed them in Figure \ref{ZTFLCs}. The maximum variability amplitude is estimated between 1.11 to 2.31 for the r-band and 0.90 to 2.32 for the g-band, suggesting a strong optical variability in these objects. \cite{2024yCat..75101791N} estimated the variability amplitude ($\psi$) of a large sample of FSRQs and BL Lacs, and they show that the $\psi$ for most of the objects lies between 0-1, which is smaller than the $\psi$ estimated for NLSy1 in our work. This again suggests that optical is more variable in these objects compared to blazars, most probably due to disk contribution in NLSy1. We also derived the color-magnitude variations for our sample, and we observed a mixed trend. In some cases, such as CGRaBS J1222+0413 and PMN J0948+0022, we observed a hint of a bluer-when-brighter (BWB) trend, whereas for 1H 0323+342 and PKS 1502+036, a RWB trend is observed. In the case of RGB J1644+263, we observed a mixed trend of both RWB and BWB (see Figure \ref{COLORS}). In \cite{2024yCat..75101791N}, authors have shown that most of the FSRQs show an RWB trend, while BL Lacs show a bluer-when-brighter (BWB) trend.

\begin{figure*}
    \centering
    \includegraphics[angle=-90, width=0.48\linewidth]{Figfs/PKS1502+036.eps}
    \includegraphics[angle=-90, width=0.48\linewidth]{Figfs/RGBJ1644+263.eps}
    \includegraphics[angle=-90, width=0.48\linewidth]{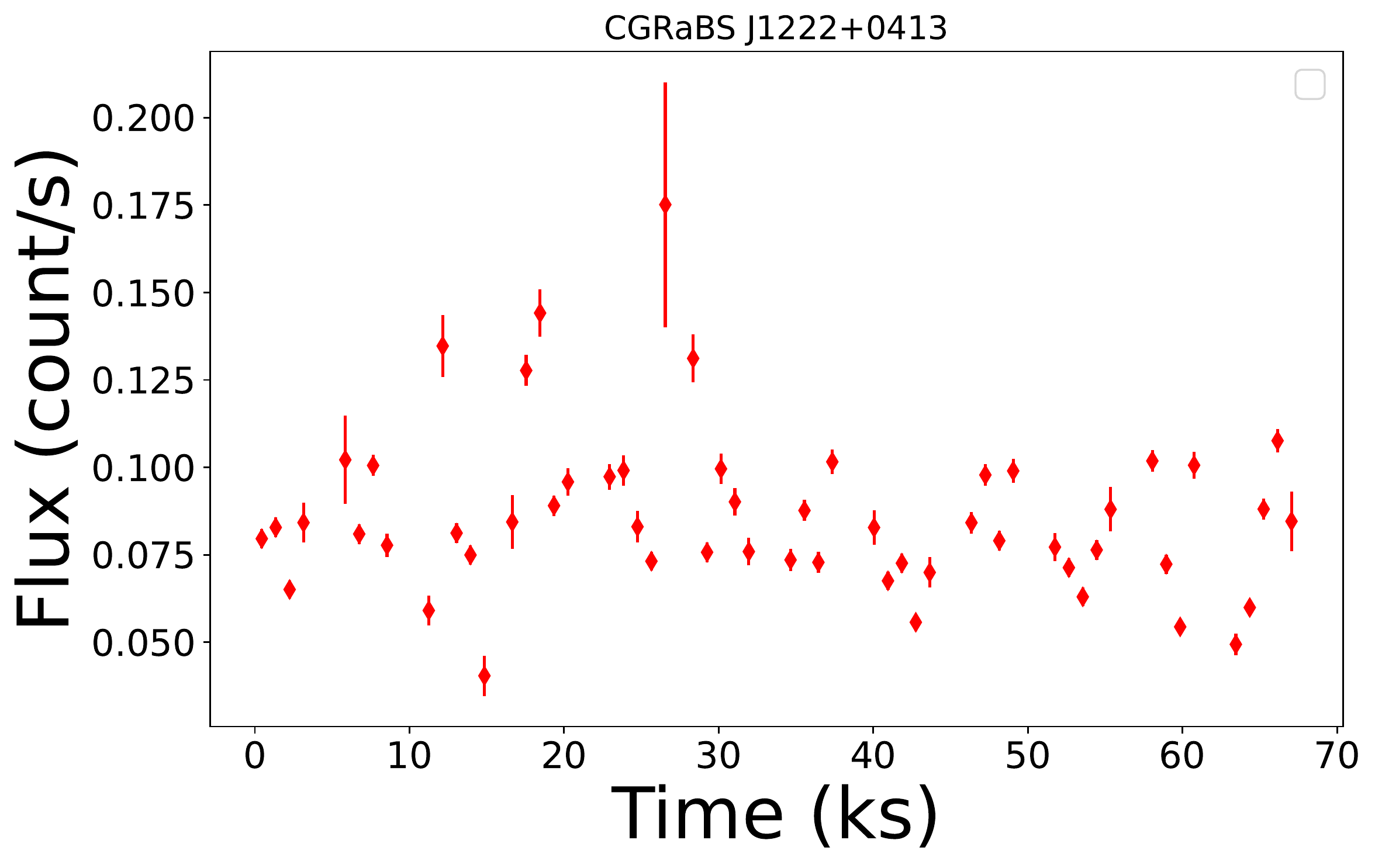}
     \includegraphics[angle=-90, width=0.48\linewidth]{Figfs/PMNJ0948+0022.eps}
    \caption{NuSTAR X-ray spectra of $\gamma$-NLSy1 fitted with power-law (PL) model.}
    \label{fig:Spectra}
\end{figure*}

\begin{figure*}
    \centering
 
    \includegraphics[angle=-90, width=0.48\linewidth]{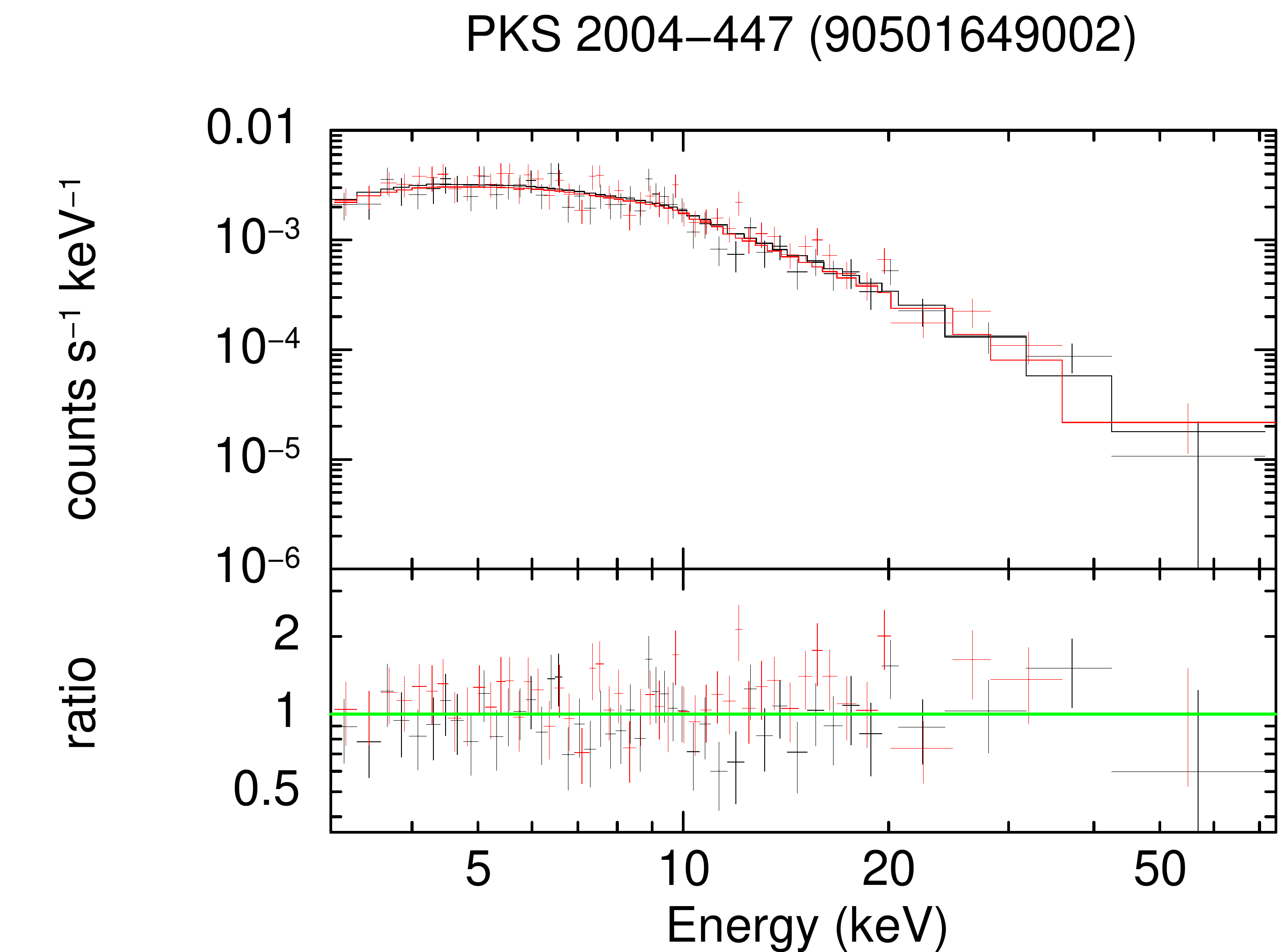}
    \includegraphics[angle=-90, width=0.48\linewidth]{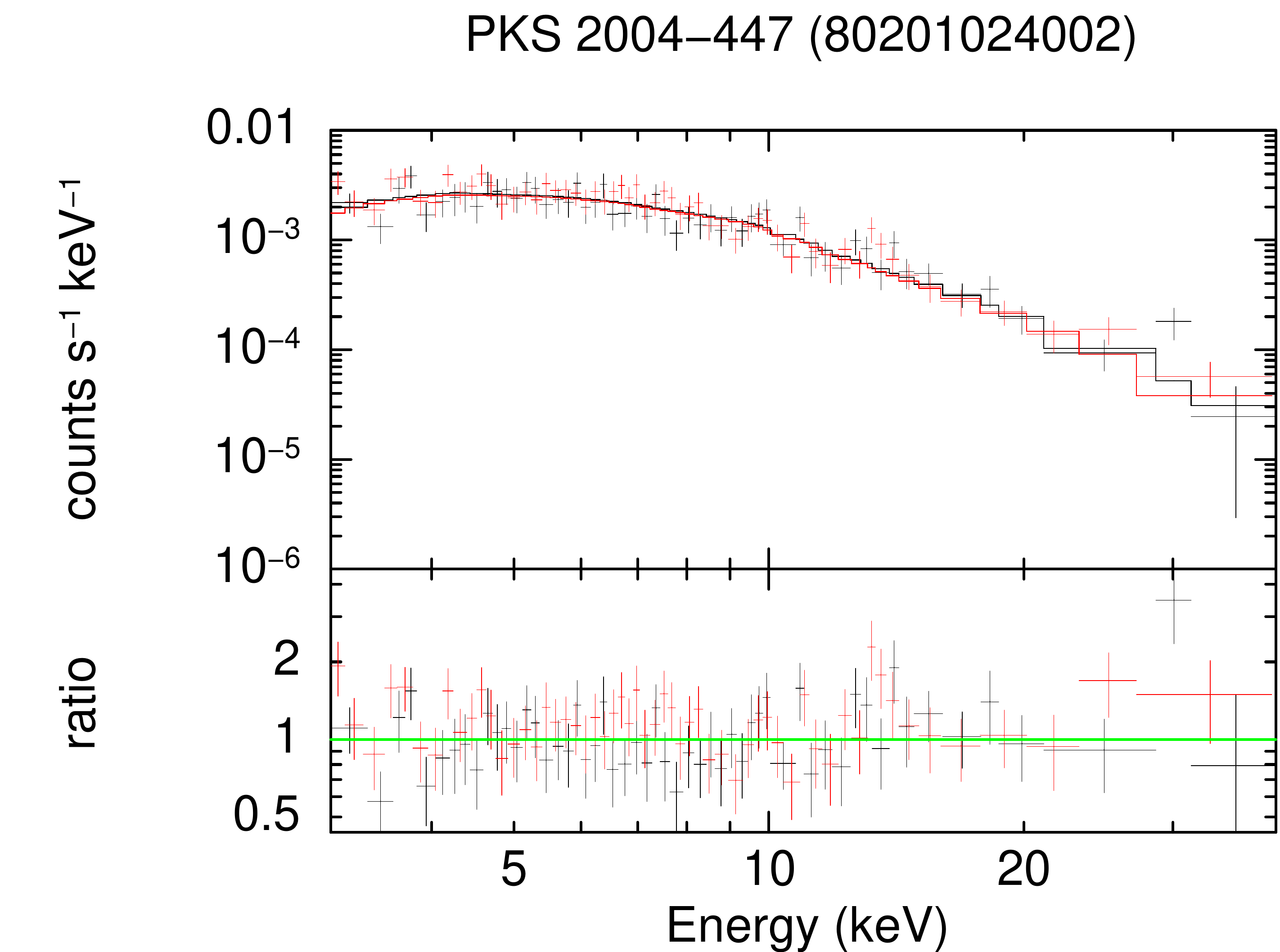}
    \includegraphics[angle=-90, width=0.48\linewidth]{Figfs/PKS2004_90501649002.eps}
    \caption{NuSTAR spectra of PKS 2004-447 fitted with a single power-law model.}
    \label{fig:enter-label}
\end{figure*}

\begin{figure*}
    \centering
    \includegraphics[angle=-90, width=0.45\linewidth]{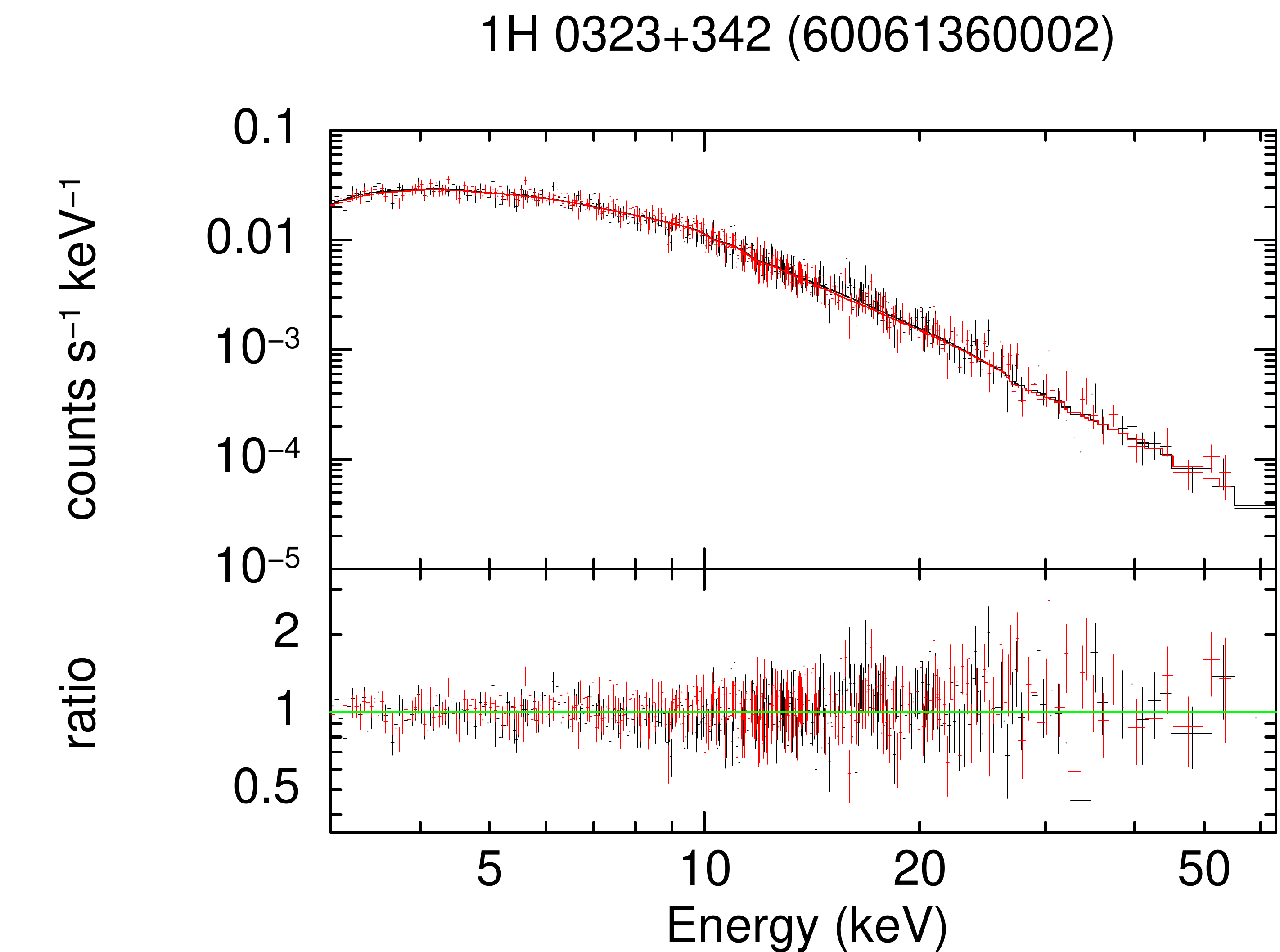}
    \includegraphics[angle=-90, width=0.45\linewidth]{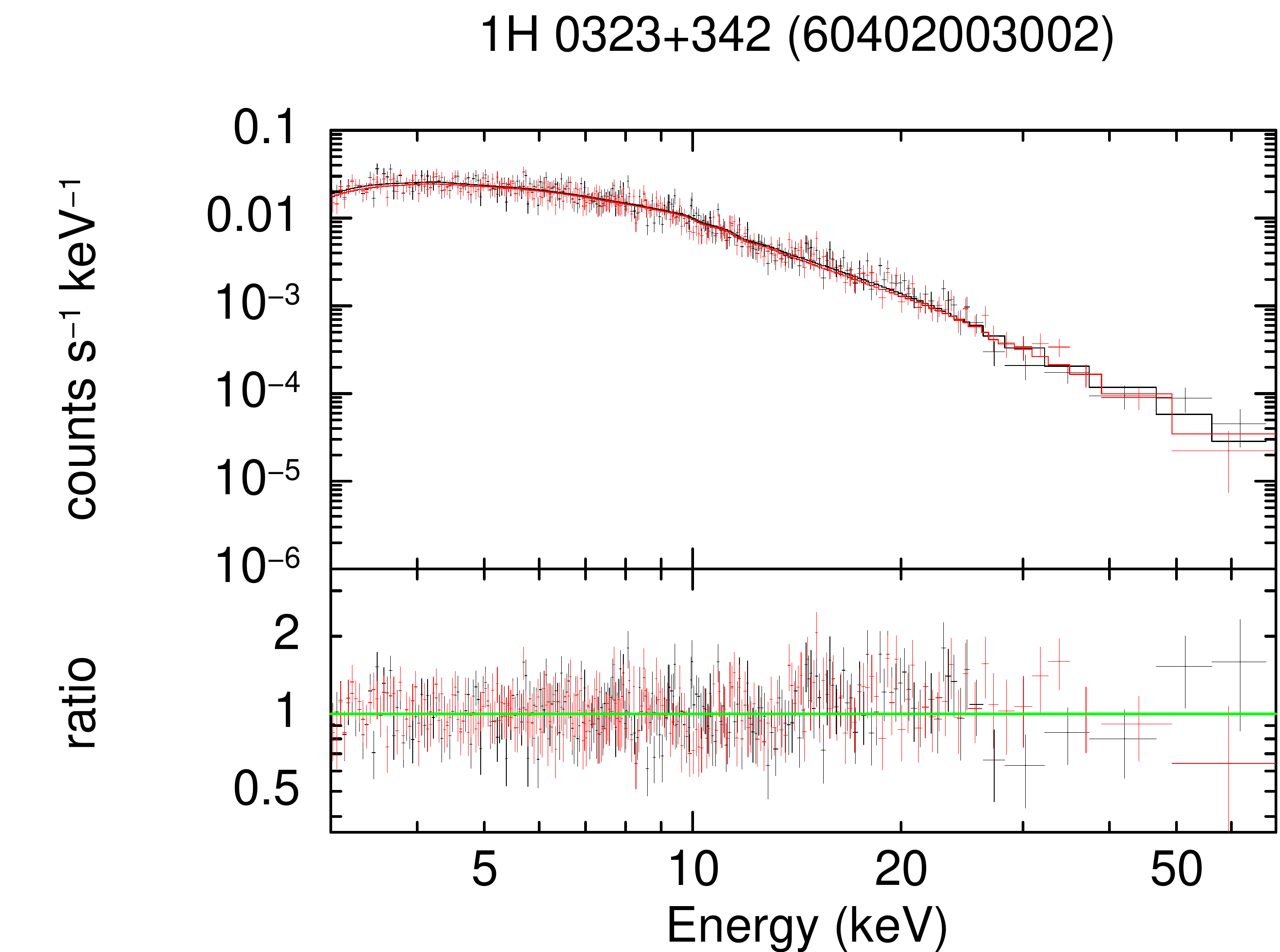}
    \includegraphics[angle=-90, width=0.45\linewidth]{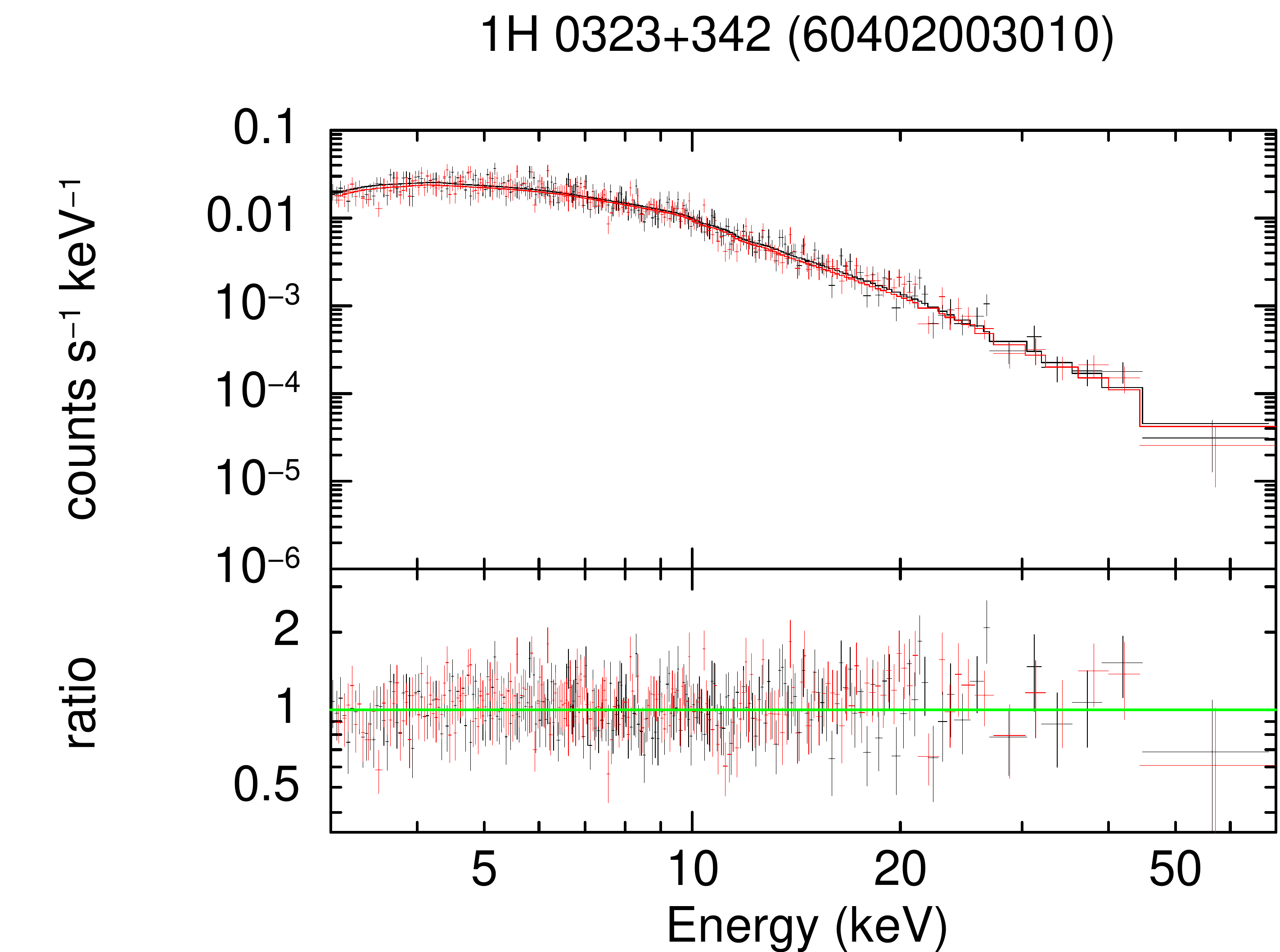}
    \includegraphics[angle=-90,width=0.45\linewidth]{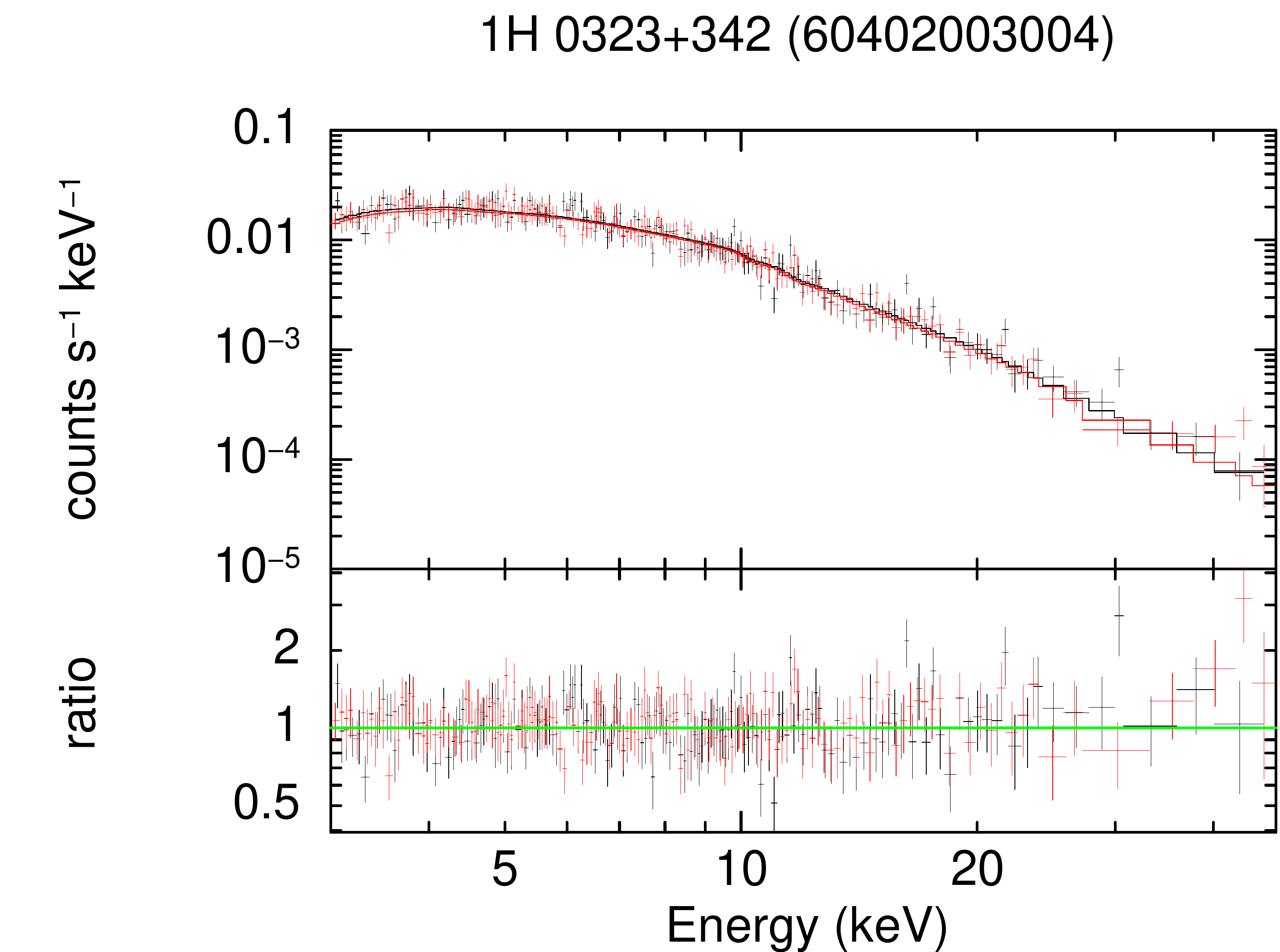}
    \includegraphics[angle=-90,width=0.45\linewidth]{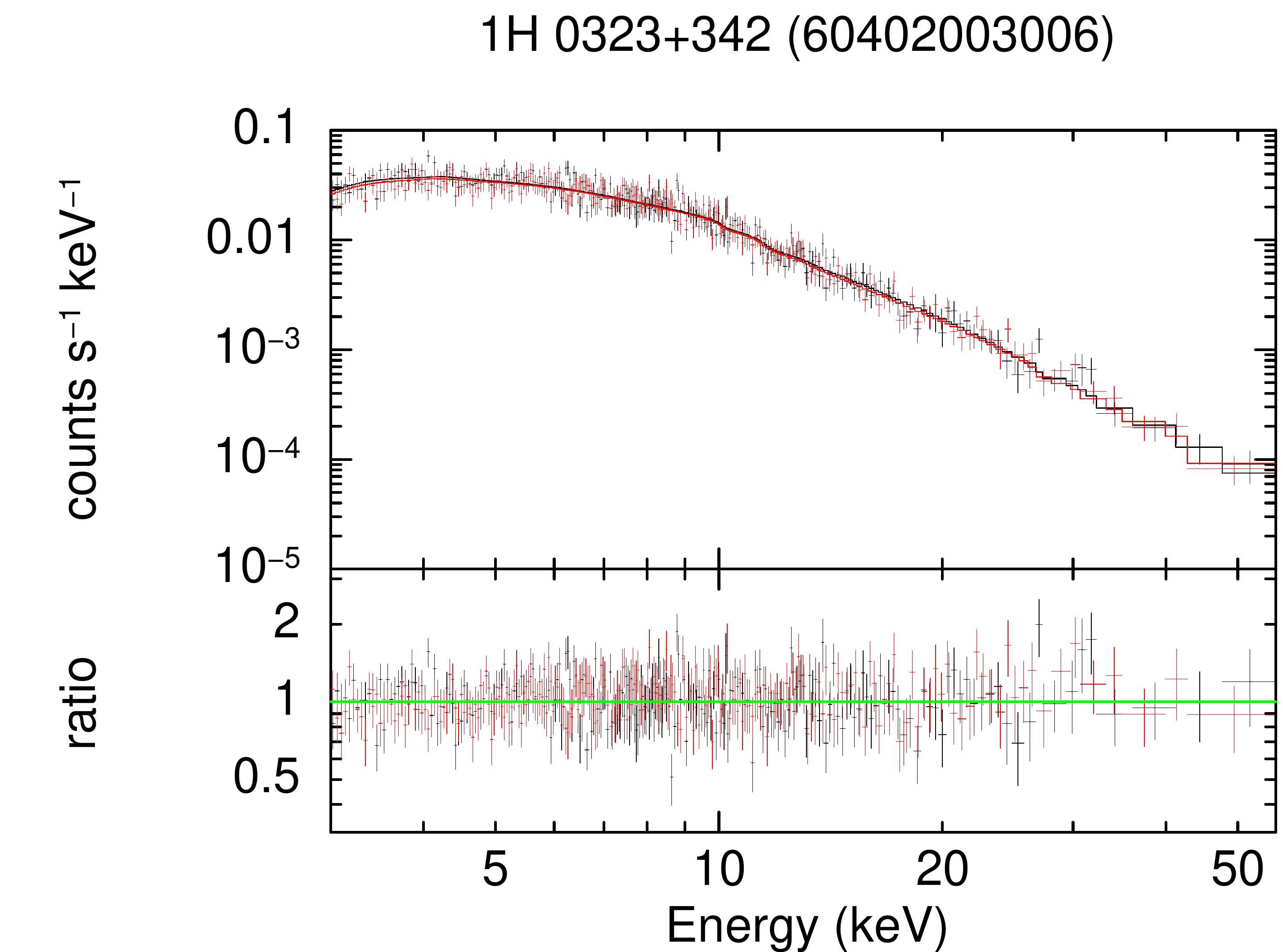}
    \includegraphics[angle=-90,width=0.45\linewidth]{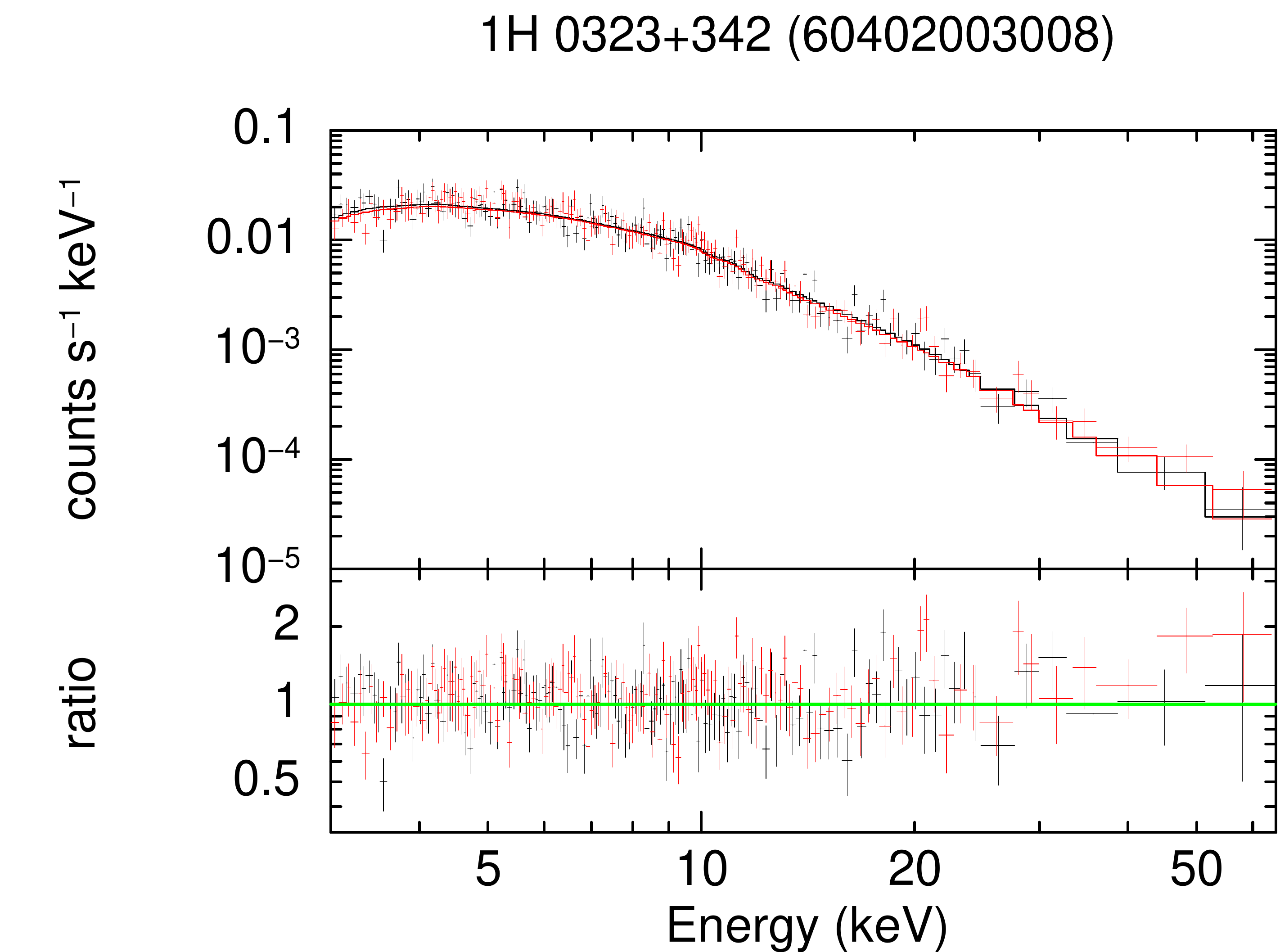}
    \includegraphics[angle=-90,width=0.45\linewidth]{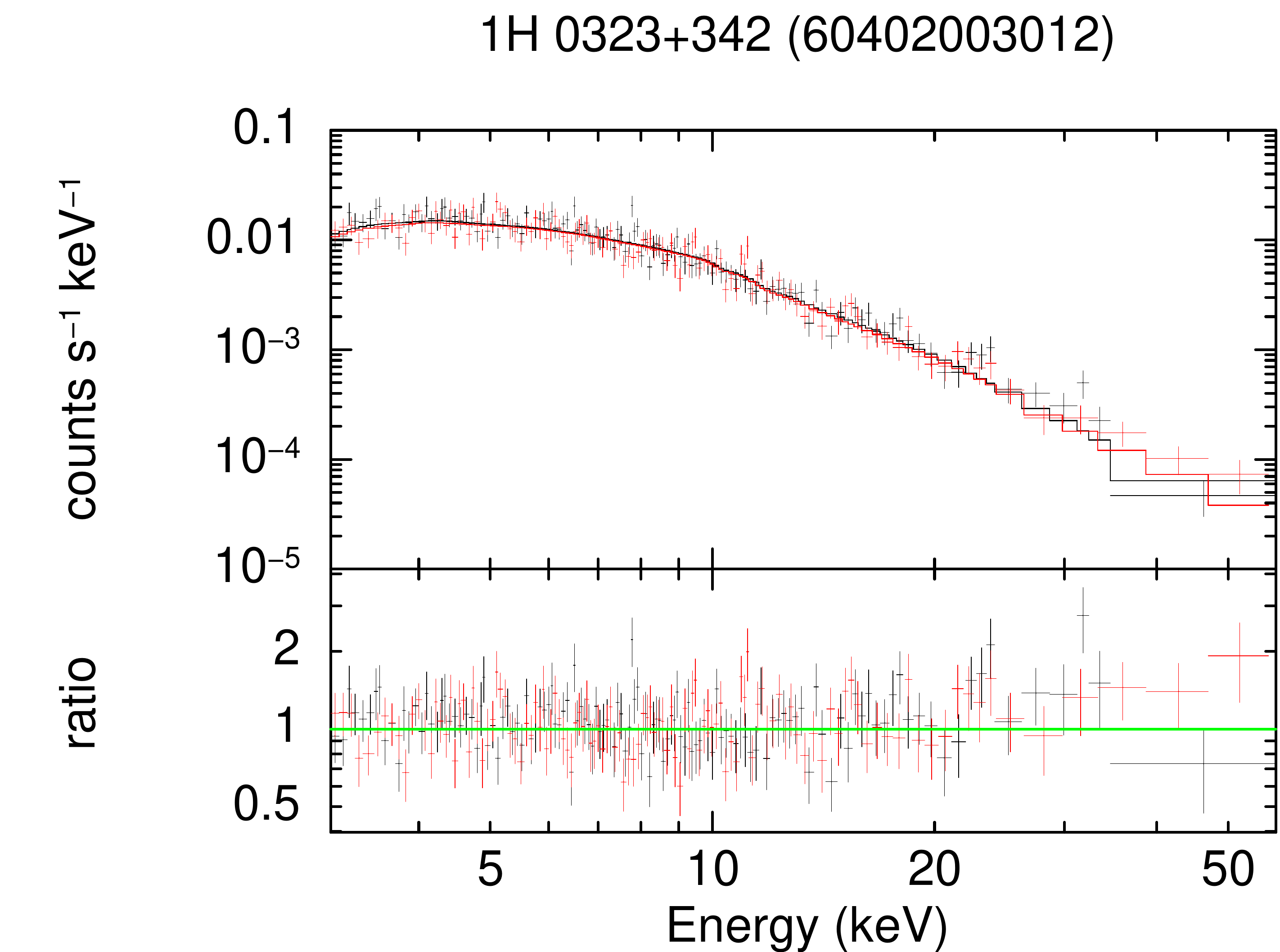}
    \includegraphics[angle=-90,width=0.45\linewidth]{91101637002.eps}
    \caption{NuSTAR spectra for various observations of 1H 0323+342.}
    \label{1Hspec}
\end{figure*}

\begin{figure*}
    \centering

   \includegraphics[angle=0, width=0.480\linewidth]{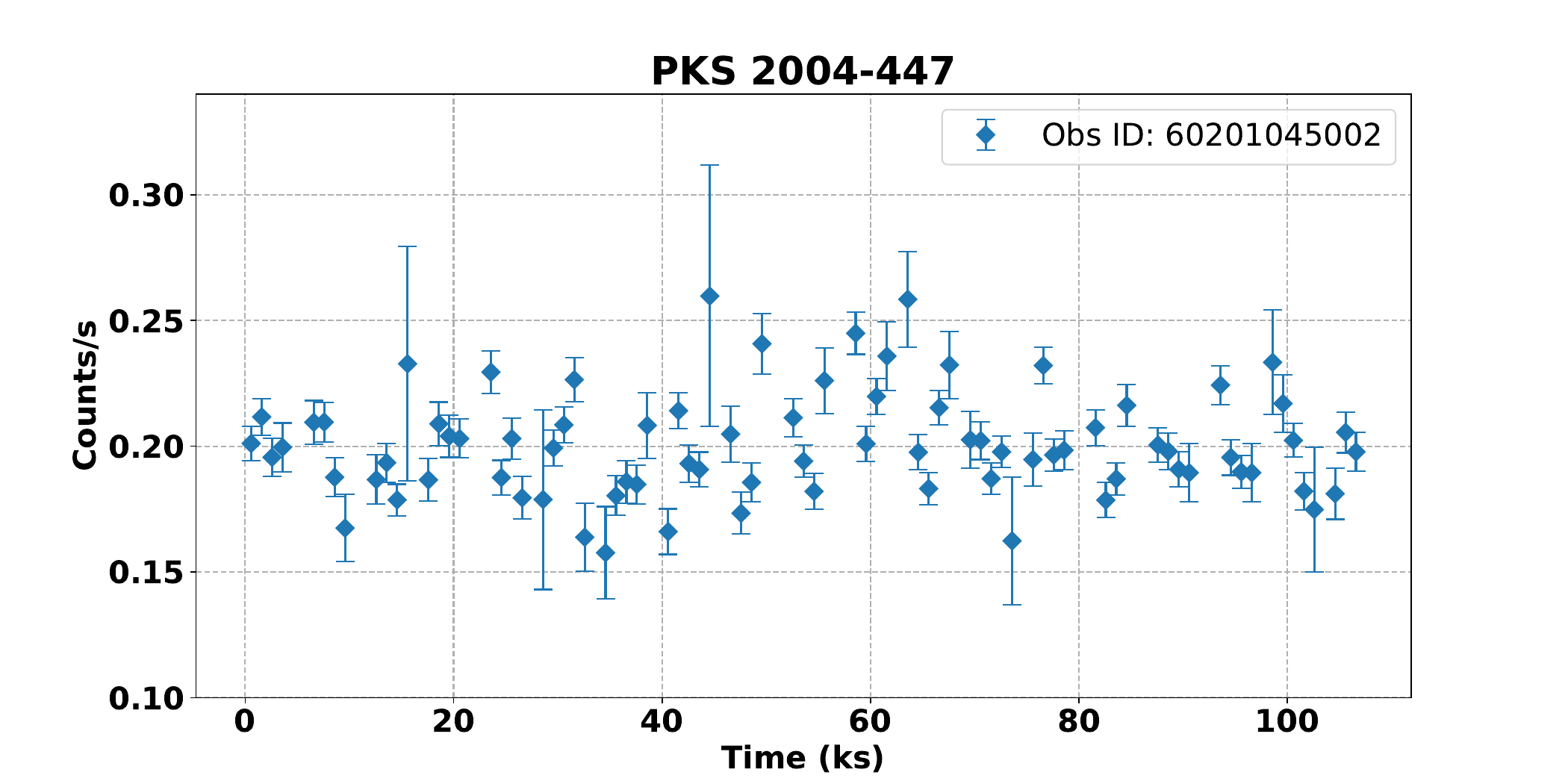}
   \includegraphics[angle=0, width=0.480\linewidth]{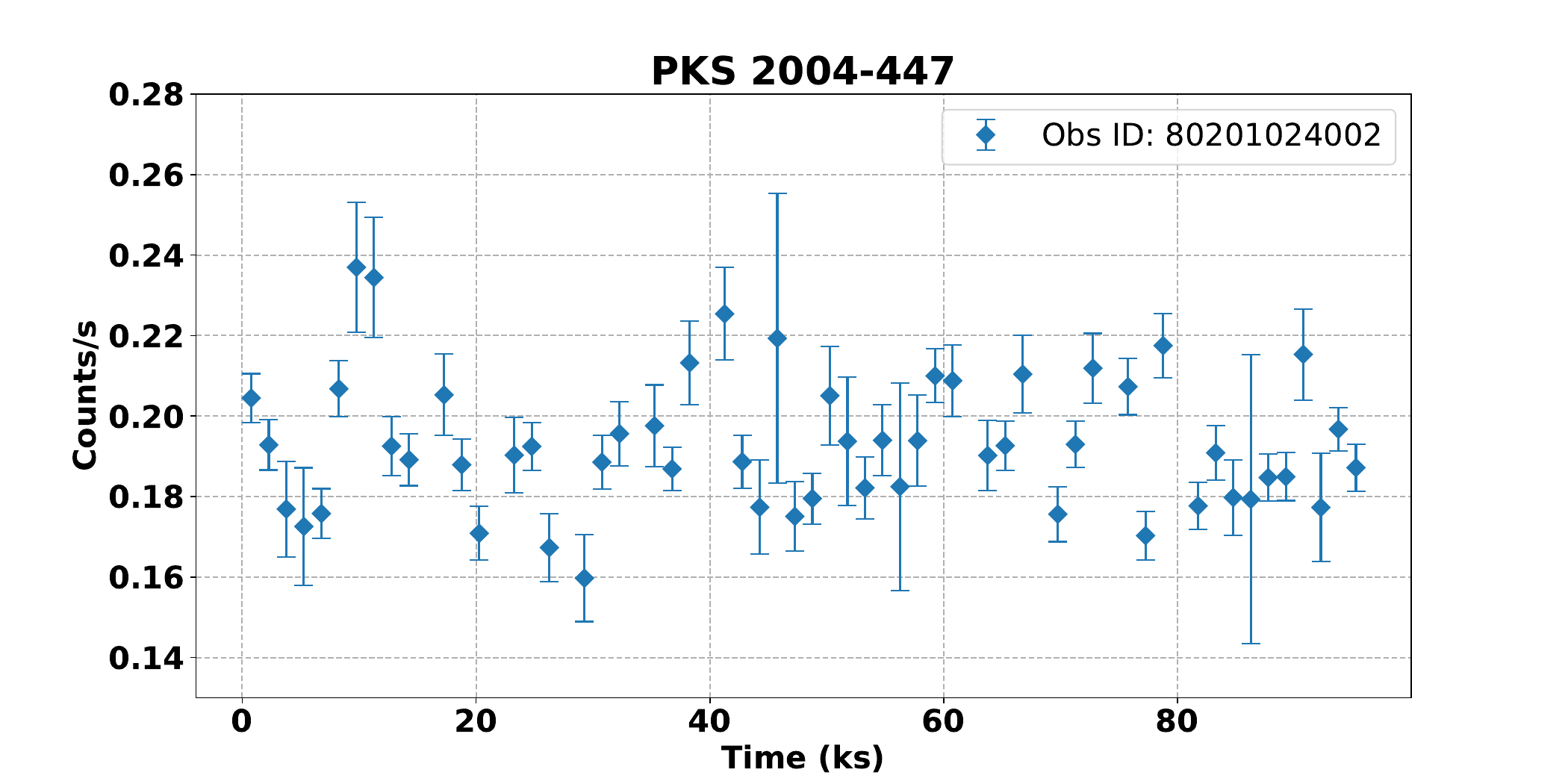}
    \includegraphics[angle=0, width=0.480\linewidth]{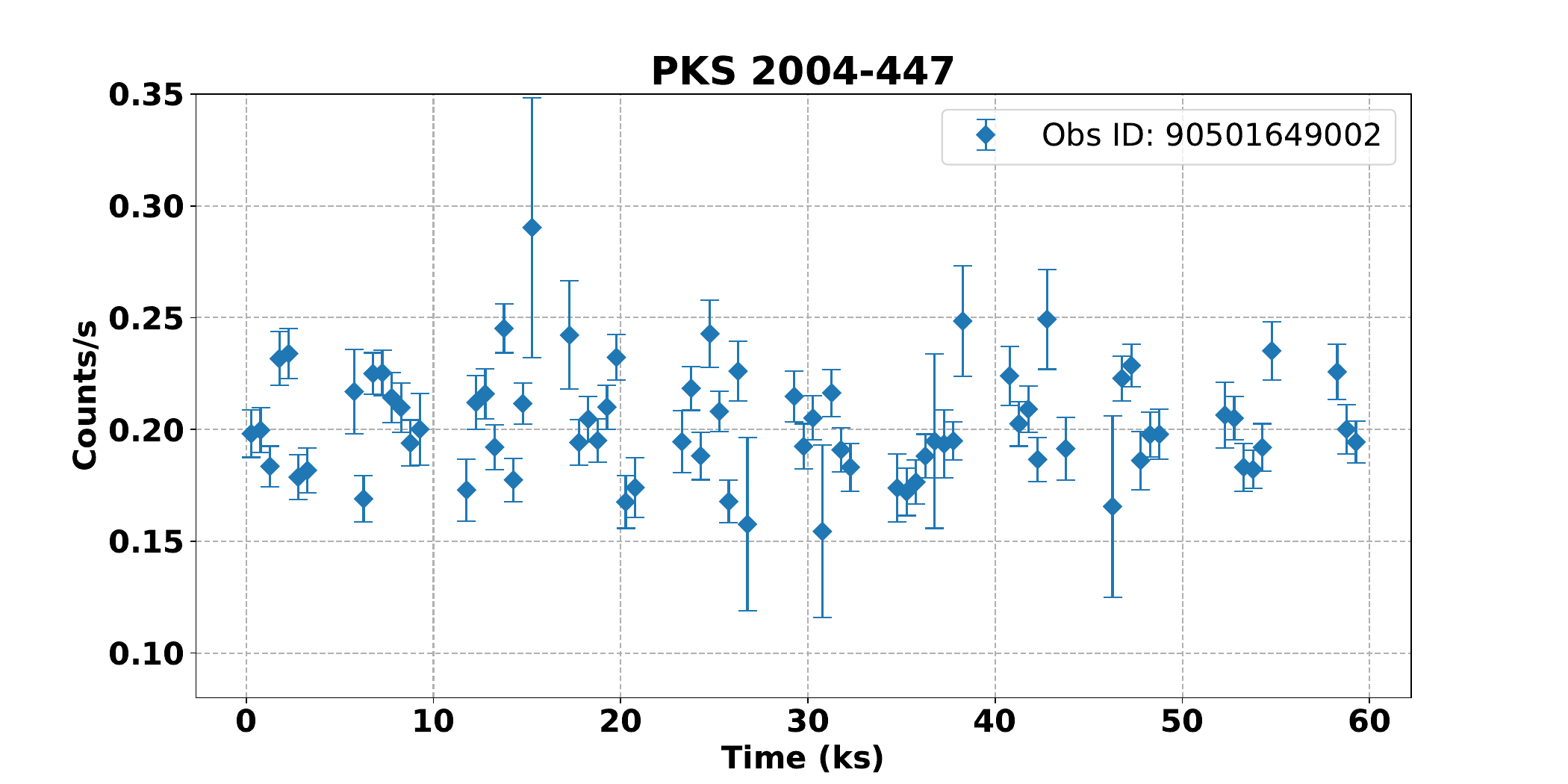}
\caption{X-ray (NuSTAR) light curve for PKS 2004-447.}
\label{LCs1}
\end{figure*}

\begin{figure*}
     \includegraphics[angle=0, width=0.480\linewidth]{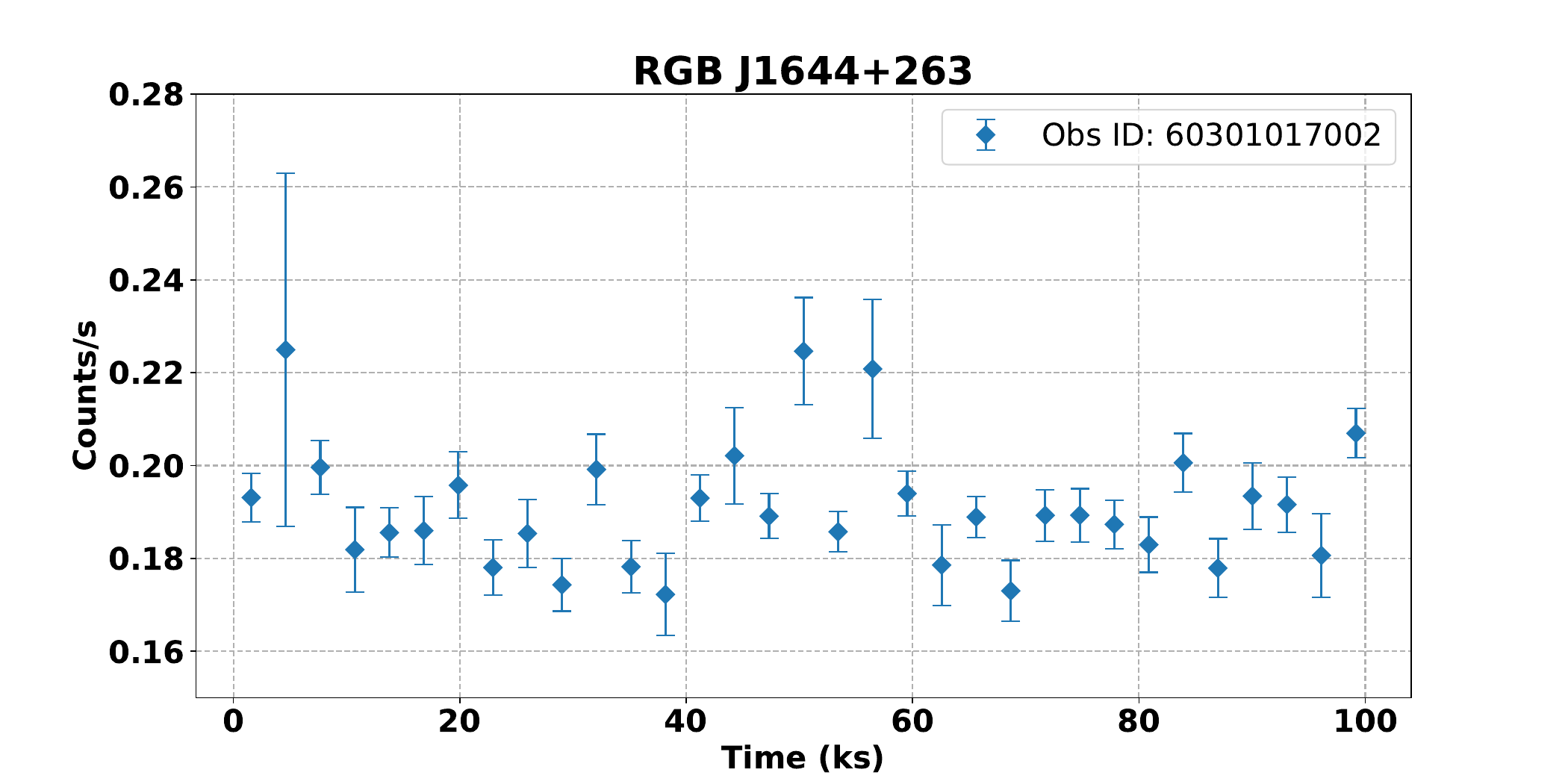}
     \includegraphics[angle=0, width=0.480\linewidth]{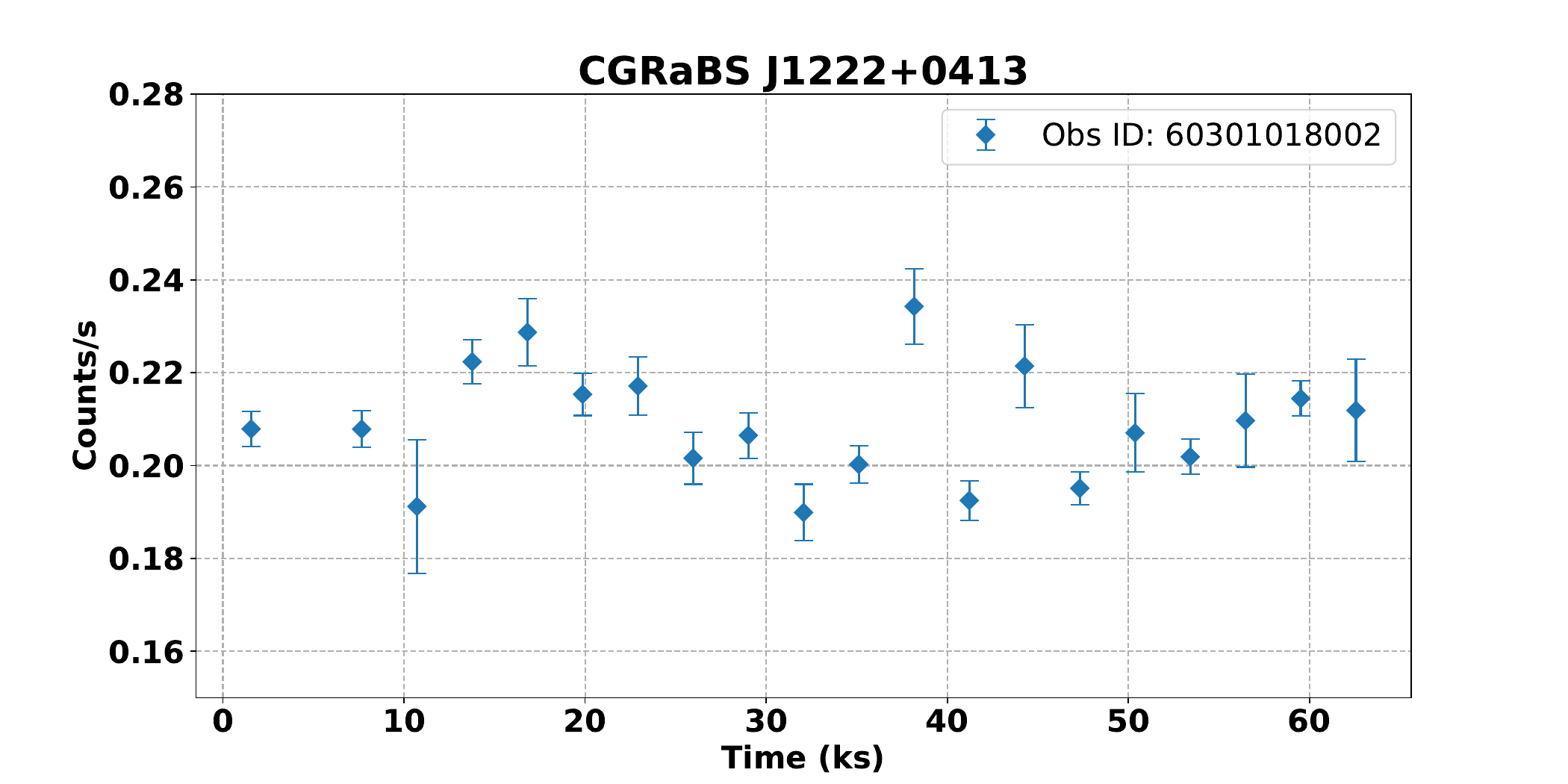}  
      \includegraphics[angle=0, width=0.480\linewidth]{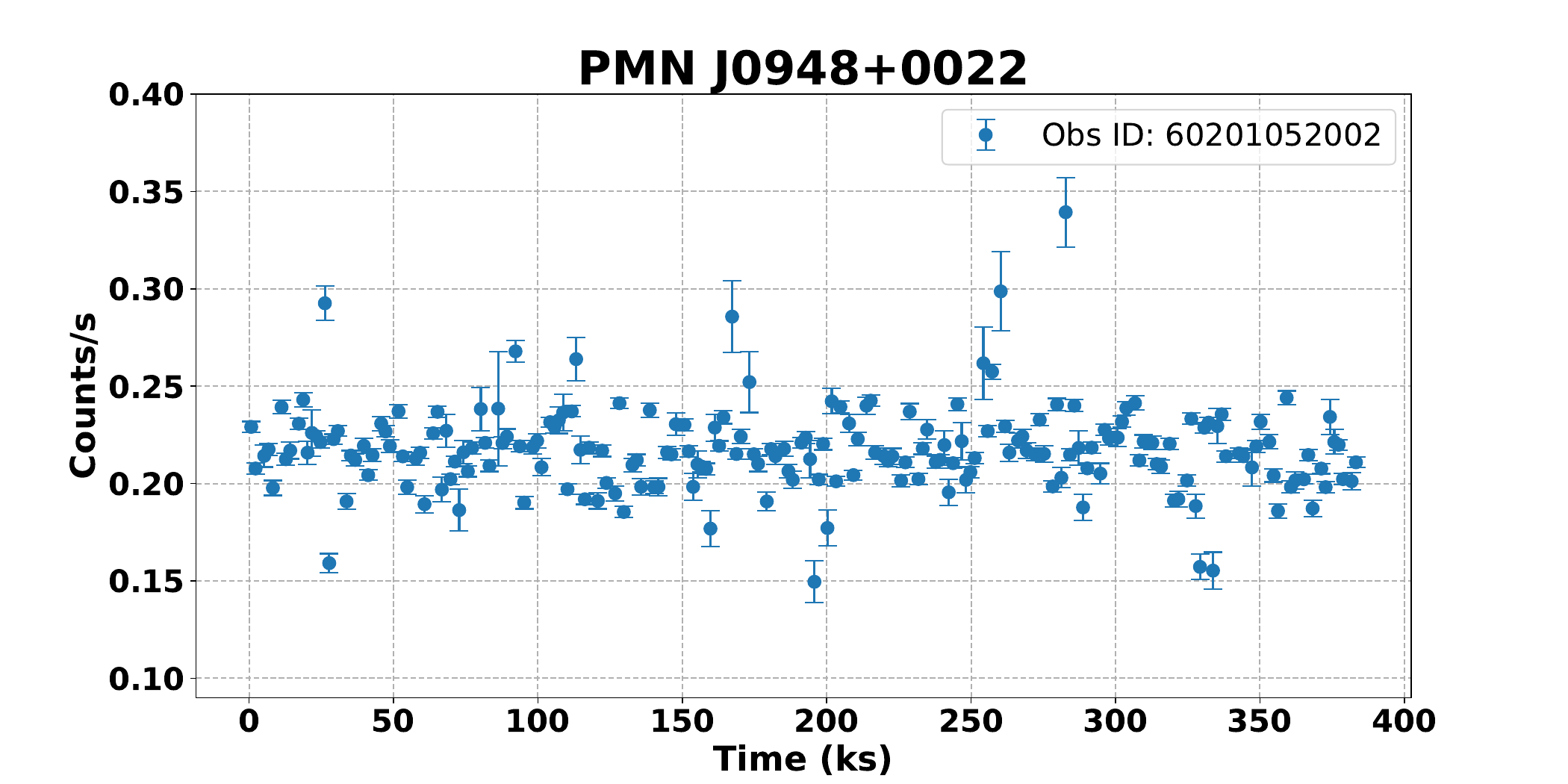}
       \includegraphics[angle=0, width=0.480\linewidth]{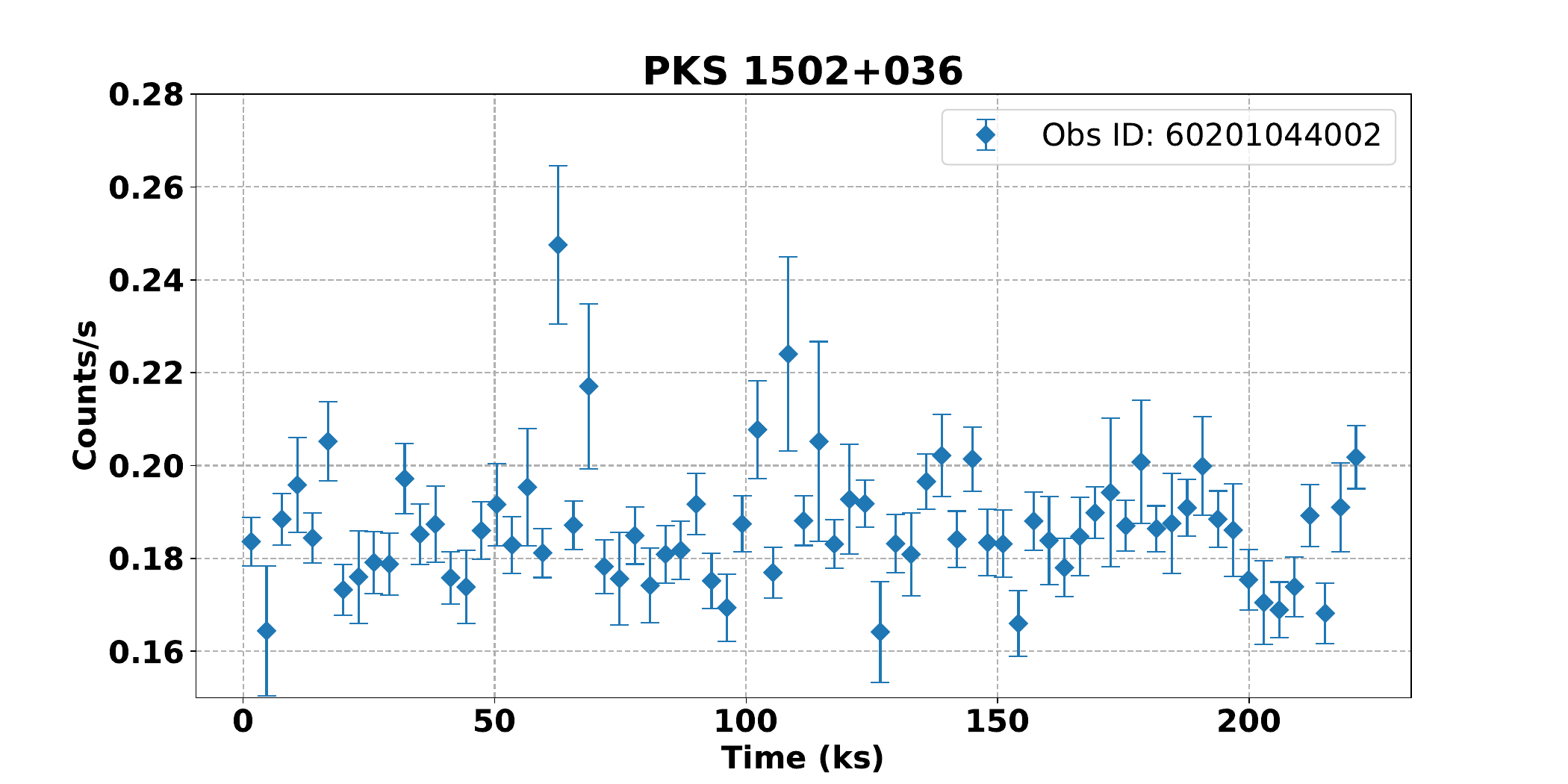}
     \caption{X-ray (NuSTAR) light curve for RGB J1644+263, CGRaBS J1222+0413, PMN J0948+0022 and PKS 1502+036.}
     \label{LCs2}
\end{figure*}

\section{CONCLUSIONS} \label{conclusion}
In this work, we have investigated 14 NuSTAR observations of six $\gamma$-NLSy1. The key highlights of this work are presented as follows:

   \begin{enumerate}
      \item The Majority of our sources exhibit a moderate hard X-ray variability with $F_{var}$ ranging 10-20\%, which is much lower than the blazars.
      \item X-ray spectra are well fitted by the power-law model, suggesting the non-thermal emission is jet-dominated.
      \item  The X-ray photon index $\Gamma_{X-ray}$  suggests that these $\gamma$-NLSy1 galaxies exhibit spectral characteristics similar to those of IBL and LBL blazars.
      \item The X-ray flux and $F_{var}$ for 1H 0323+342 show a mild anti-correlation, but there is no discernible pattern between $\Gamma_{X-ray}$ and $F_{var}$.
      \item A tight anti-correlation between $P_{jet}$ and X-ray flux strongly indicates that the X-ray emission is jet-dominated, suggesting that jet activity affects high-energy radiation, which corresponds to AGN jet models where synchrotron or inverse Compton processes within the jet emit X-rays.
      \item We have observed a moderate anti-correlation between $L_{disk}$ with X-ray flux, which implies that when the disk gets more luminous, the X-ray flux goes down, indicating variations in the accretion flow or disk dissipation.
      \item The optical lightcurves in the g- and r-band also show a rapid flux variation on intraday and month timescales, implying a combination of jet and disk activity.
      \item A color-magnitude plot (g-r vs r) shows a mixed trend of BWB and RWB, suggesting that jet emission pushes the spectrum toward the blue or red end as it rises, beyond the accretion disk's thermal contribution. This trend is consistent with other jetted AGNs, like blazars.
   \end{enumerate}

\begin{acknowledgments}
This research utilizes publicly available data on various NLSy1 obtained from the HEASARC webpage and the ZTF archive. RP acknowledges the support of the IoE seed grant from BHU.
\end{acknowledgments}

\bibliography{sample7}{}
\bibliographystyle{aasjournalv7}

\end{document}